\documentclass[twocolumn,prb,amsmath,amssymb,floatfix,superscriptaddress]{revtex4-2} 
\usepackage{amsmath}
 \usepackage{gensymb} 
\usepackage{amssymb}
\usepackage{graphicx}
\usepackage{bm}
\usepackage{color}
\usepackage{hyperref}
\usepackage{graphicx}
\usepackage{upgreek}
\usepackage{scalerel}
\usepackage{multirow}
 
 \usepackage[bb=dsserif]{mathalpha}
 
\usepackage{pgffor}
\usepackage{pdfpages}
\usepackage{mathdots}
\usepackage{float}
\usepackage{silence}
\usepackage{physics} 
\usepackage{lscape}   
\usepackage{color, colortbl}
\usepackage[table]{xcolor}
\usepackage{comment}  

\usepackage{soul}

\hypersetup{ colorlinks, citecolor=blue, linkcolor=blue, urlcolor=blue}

\makeatletter
\AtBeginDocument{\let\LS@rot\@undefined}
\makeatother

\begin{document}
 \title{Josephson diode effect with Andreev and Majorana bound states}

\author{Sayan Mondal}
 \affiliation{Department of Physics and Astronomy, Uppsala University, Box 516, S-751 20 Uppsala, Sweden}
	
 \author{Pei-Hao Fu}
 \affiliation{School of Physics and Materials Science, Guangzhou University, Guangzhou 510006, China}
	
 \author{Jorge Cayao}
 \affiliation{Department of Physics and Astronomy, Uppsala University, Box 516, S-751 20 Uppsala, Sweden}
\date{\today}	
	
\begin{abstract}
Superconductor-semiconductor hybrids have been shown to be useful for realizing the Josephson diode effect, where nonreciprocity in the supercurrents occurs due to the interplay of the Josephson effect and applied magnetic fields.  With the same ingredients, these Josephson junctions can also host Andreev and Majorana bound states, whose interplay with the Josephson diode effect is however not fully understood. In this work, we consider short Josephson junctions based on superconductor-semiconductor systems under homogeneous Zeeman fields  and investigate the Josephson diode effect in the presence of Andreev and Majorana states. Under generic conditions, the  Zeeman field component parallel to the spin-orbit axis   promotes an asymmetric low-energy spectrum as a function of the superconducting phase, which persists in the trivial and topological phases hosting Andreev and Majorana bound states, respectively. Interestingly, this spectrum asymmetry originates   supercurrents that are not odd functions of the superconducting phase difference as in common Josephson junctions, thereby developing a nonreciprocal behaviour that signals the emergence of the Josephson diode effect. We show that the Josephson diode effect  is particularly promoted under the presence of both zero-energy Andreev and Majorana bound states, revealing that Josephson diodes can be realized in the trivial and topological phases of superconductor-semiconductor hybrids. We then demonstrate that  the Zeeman field evolution of the diode's efficiencies is able to map the topological phase transition and the formation of Majorana bound states via an oscillatory behavior that becomes more visible in long superconductors.  While Josephson diodes generally exist in the trivial and topological phases of Josephson junctions, we discover that in the tunneling regime only a Josephson diode effect in the topological phase remains due to the finite contribution of Majorana bound states. Our findings help understand the Josephson diode effect in superconductor-semiconductor hybrids and can also be useful for guiding the realization of Majorana-only Josephson diodes as well as for identifying Majorana states.
\end{abstract}
	
\maketitle

\section{Introduction}
 Josephson junctions (JJs)  have been one of the most studied systems in condensed matter physics  not only because they can host novel   physics but also due to their promising applications \cite{josephson1962,likharev1979,RevModPhys.73.357,golubov2004,RevModPhys.77.935,BergeretReview,birge2024ferromagnetic,RevModPhys.96.021003,acin2018quantum,fukayaCayaoReview2025_IOP}. JJs composed of two coupled superconductors enable the flow of a dissipationless supercurrent known as the Josephson effect \cite{josephson1962},  which is carried by Andreev bound states   due to a finite phase difference between superconducting order parameters \cite{kulik1975,furusaki1991dc,PhysRevB.45.10563,beenakker1992,Furusaki_1999,kashiwaya2000,PhysRevB.64.224515,PhysRevLett.96.097007,golubov2004,sauls2018,mizushima2018}.  The Josephson effect (JE) and Andreev bound states (ABSs) in JJs have been shown to be   crucial for   superconducting qubits \cite{devoret2005implementing,wendin2007quantum,clarke2008superconducting,kjaergaard2020,aguado2020perspective,aguado2020majorana,benito2020hybrid,siddiqi2021engineering}, superconducting spintronics 
\cite{eschrig2011spin,linder_2015,Eschrig2015,yang2021boosting,mel2022superconducting,cai_2023},   superconducting quantum interference devices  \cite{jaklevic1964, silver1967, kleiner2004,clarke2006squid,granata2016nano,seredinski2019quantum}, and, more recently,   also for realizing Josephson diodes \cite{Jiangping07,PhysRevB.92.035428,PhysRevB.93.174502,Misaki21,tanaka2022,davydova2022,nadeem2023superconducting}. In this regard, Josephson diodes (JDs) are of particular relevance because they result from nonreciprocal supercurrents \cite{Jiangping07,PhysRevB.92.035428,PhysRevB.93.174502,Misaki21,tanaka2022,davydova2022,nadeem2023superconducting,maian2023, souto2022, costa2023, davydova2022,  debnath2024, scharf2024, correa2024, shen2024, soori2025, nikodem2024, debnath2024field,mazur2024, wu2022, baumgartner2022, baumgartner2022_2, pal2022, kudriashov2025,hu2023,zhang2022,ding2024,Turini2022,Qiang24,banerjee2024_2,kotetes2024,trahms2023diode,ilic2022,liu2024, souto2024, fracassi2024,mazur2024,legg2023,karabassov2022, fu2024, legg2023, lu2023, cuozzo2024,cayao2024_JD,meyer2024josephson,valentini2023parityconserving,hou2023,aligia2020,Lotfizadeh_2024,yerin2024,maiellaro2024}, which makes them  useful for dissipationless, or low-dissipation, circuit elements in superconducting devices \cite{braginski2019superconductor,gallop2017squids,anders2010european,hoshino2018, wakatsuki2017,nagaosa2024nonreciprocal,tokura2018nonreciprocal,zapata1996,beck2005}. JDs can thus pave the way for transformative advancements in electronic device technologies \cite{nadeem2023superconducting,sze2008semiconductor,mehdi2017thz,semple2017flexible,coldren2012diode}.

It is by now understood that the necessary conditions for realizing JDs involve breaking time-reversal and inversion symmetries \cite{tanaka2022,zhang2022, yuan2022, daido2022,he2024,FAndoNature2020}. While these requirements can be achieved in distinct setups \cite{nadeem2023superconducting,fukayaCayaoReview2025_IOP},   superconductor-semiconductor hybrids under magnetic fields have  attracted considerable attention \cite{davydova2022,ilic2022,liu2024, souto2024, fracassi2024,mazur2024,legg2023,karabassov2022, tanaka2022, fu2024, legg2023, lu2023, cuozzo2024,cayao2024_JD,meyer2024josephson,PhysRevB.93.174502,PhysRevB.92.035428,valentini2023parityconserving,hou2023,baumgartner2022,baumgartner2022_2,he2024,aligia2020,Lotfizadeh_2024} due to their promising properties and  great experimental and theoretical advances \cite{sato2016,Aguadoreview17,sato2017,lutchyn2018,prada2020,Cayao2020odd,flensberg2021,frolov2020,marra2022,tanaka2024,tanaka2012}. In fact,  JJs  based on  superconductor-semiconductor hybrids  are predicted to host a topological phase characterized by the emergence of four Majorana bound states (MBSs) at large Zeeman fields \cite{sanjose2012,jose2013,cayao2015,cayao2017,peng2016,cayao2018,cayao2021,perkerten2022,baldo2023,awoga2019}, in addition to the ABSs present in the trivial phase 
\cite{beenakker1992,sauls2018,mizushima2018,hays2018, tosi2019, ren2019, nichele2020}; see also Refs.\,\cite{sato2016,sato2017,Aguadoreview17,lutchyn2018,prada2020,Cayao2020odd,flensberg2021,frolov2020,marra2022,tanaka2024}. While JDs have  been widely studied in the trivial phase with ABSs, not many studies addressed JDs with MBSs. In particular, there exist limited works addressing JDs with four MBSs in    superconductor-semiconductor hybrids \cite{cayao2024_JD,liu2024}, where  they consider transparent JJs with an inhomogeneous magnetic field that does not affect the topological protection of MBSs but  is rather challenging to achieve. In this regard, the effect of  MBSs on the efficiency of JDs when the topological protection is reduced as well as the possibility to promote JDs only with MBSs remain to be addressed.

In this work, we consider JJs with Rashba spin-orbit coupling (SOC), which can be  realized in superconductor-semiconductor hybrids, and investigate the emergence of JDs when   homogeneous magnetic fields are applied. We find that, when the Zeeman field has a component parallel   to the SOC, the Andreev spectrum is asymmetric with respect to the superconducting phase difference, including regimes with ABSs and MBSs in the trivial and topological phases of transparent JJs, respectively.  We then demonstrate that this asymmetry induces current-phase curves possessing distinct positive and negative critical currents which gives rise to nonreciprocal Josephson transport and to   JDs. We obtain that, while the efficiency of the JDs is largest in the trivial regime of transparent JJs, the lower efficiencies of JDs in the topological phase trace the topological phase transition and the formation of MBSs. These signatures can be proved by the Zeeman dependence of the diode's efficiencies, which    develop  a kink at the topological phase transition and an oscillatory pattern  uniquely associated to the spatial nonlocality of MBSs. By increasing the length of the superconductors, the efficiency of the JDs in the topological phase  reaches higher values and its oscillations develop  short periodicity as a response to MBSs becoming more zero energy and spatially nonlocal. We further discover that, while JDs emerge in the presence of   ABSs and MBSs in transparent JJs, by reducing the junction's transmission, it is possible to lower the diode's efficiency in the trivial regime such that JDs only   emerge in the topological phase with MBSs. These results therefore provide fundamental understanding of JDs in semiconductor-based JJs with Andreev and Majorana states.
 
This paper is organized as follows: In Sec.~\ref{sec:ham}, we describe the Hamiltonian of the JJ and its band structure. Sec.~\ref{sec:abs} explores the phase- and Zeeman-field-dependent ABSs and MBSs. In Sec.~\ref{sec:jc}, we analyze the asymmetric Josephson current across the JJ. The emergence of the JD is discussed in Sec.~\ref{sec:efficiency} by examining the critical current and efficiency as a function of the Zeeman field. Finally, in Sec.~\ref{sec:conclusion}, we summarize our findings. Further,
in App.\,\ref{app:JD_with_long_N}, we present the JJ with a long normal region and discuss possible scenarios in multichannel junctions. App.\,\ref{app:JC_with_long_S} shows the low-energy spectrum and Josephson current for a JD with extended superconducting regions, while App.\,\ref{app:Localization_MBS_in_JD} addresses the localization of MBSs. Finally, in App.\,\ref{app:Efficiency_temperature}, we demonstrate the robustness of the JD’s efficiency at finite temperatures.

\begin{figure}[!t]
\centering
 \includegraphics[width=\columnwidth]{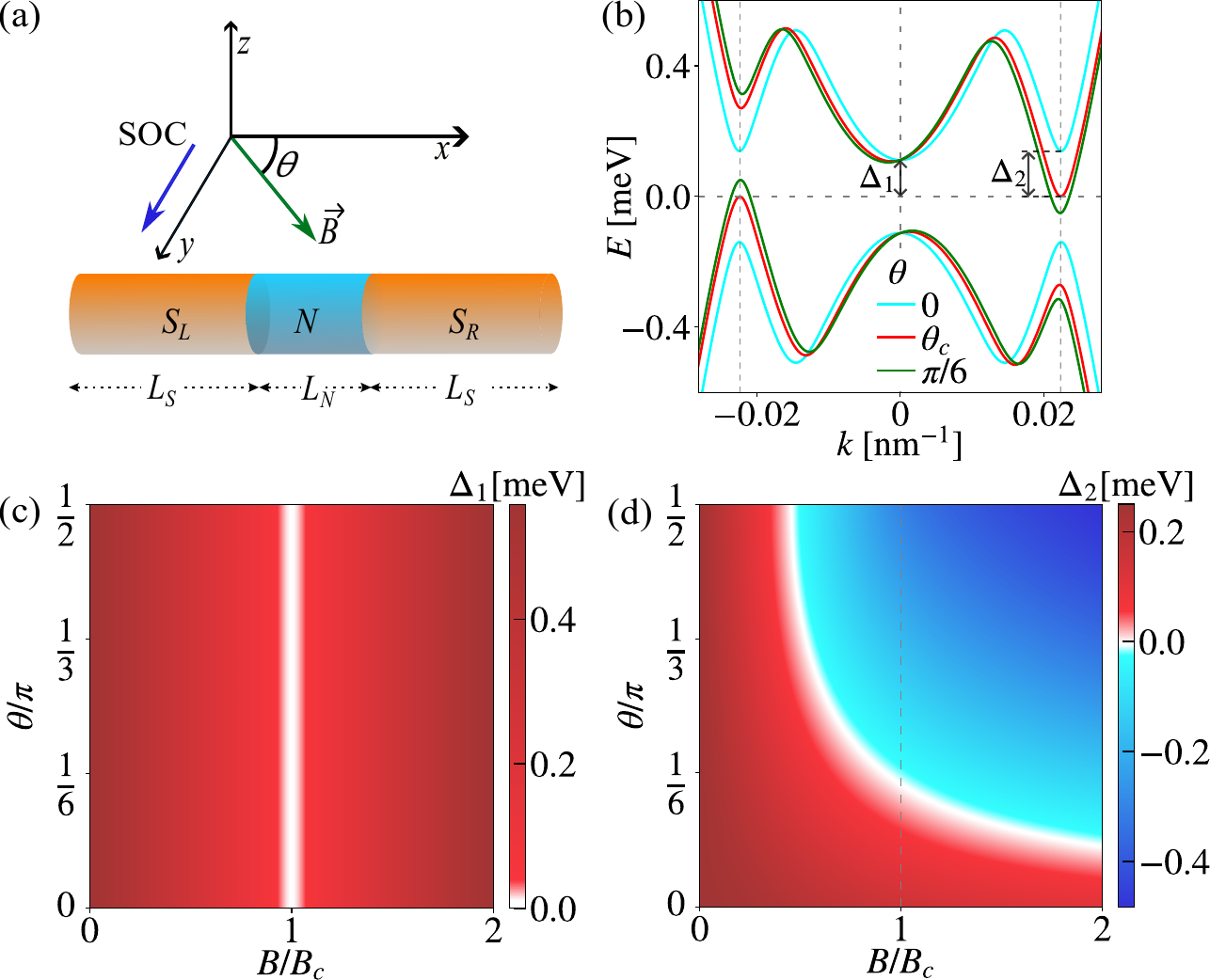}
 \caption{(a) A JJ formed by a nanowire with SOC and an homogeneous magnetic field. The left and right sides of the nanowire denoted by S$_{\rm L,R}$ (orange regions) are of finite length and contain proximity induced superconductivity from  conventional superconductors, while the finite middle light blue region  is left without superconductivity and denoted by N. The JJ is formed along $x$, while the SO axis    lies along the $y$-direction and is indicated by the blue arrow.  The Zeeman field $B$ results from an applied magnetic field  throughout the JJ and makes an angle $\theta$ with the $x$-axis, thus containing components that are parallel and perpendicular to the SO axis. (b) Positive and negative energies closest to zero versus momentum for distinct values of $\theta$, indicating  $\Delta_{1(2)}$ at zero momentum (positive Fermi momenta $+k_F$). Here, the Zeeman field   is   $B=1.2B_c$, for which  $\Delta_2=0$  at $\theta_c \approx 0.1216\pi$ indicated by red curve.  (c,d) Gaps $\Delta_{1,2}$ as a function of $\theta$ and $B$, with $\Delta_{2}$ at $+k_F$ with respect to the Fermi level. Here,  $B=B_c$ marks the topological phase transition   $\Delta_1=0$. Parameters: $\alpha_R = 20$meVnm, $\Delta = 0.25$meV, and $\mu = 0.5$meV.}
\label{Fig1} 
\end{figure}

\section{The Josephson junction model}
\label{sec:ham}
We consider a JJ based on a single-channel semiconducting nanowire with proximity induced superconductivity and an homogeneous magnetic field, see Fig.\,\ref{Fig1}(a). In momentum space, we model the semiconductor with proximity-induced superconductivity   by a Bogoliubov-de Gennes (BdG) Hamiltonian that reads \cite{sat09,sat10,lutchyn2010,oreg2010}
\begin{equation}
\label{eq:ham_BDG}
\begin{split}
H_{\rm BdG} &= \left(-\frac{\hbar^{2}\partial_{x}^{2}}{2m^{*}}-\mu\right) \tau_z \sigma_0 - i \alpha_R \tau_z \sigma_y \partial_x\\
& + B_x \tau_z \sigma_x + B_y \tau_0 \sigma_y + \Delta \tau_y \sigma_y\,,
\end{split}
 \end{equation}
 where  $m^*$ represents the electron's effective mass and $\mu$ the chemical potential measured from the bottom of the band. The second term describes the Rashba SOC with the spin-orbit (SO) axis along $y$ and $\alpha_R$ being the strength of the coupling; here $\hbar$ does not appear explicitly as $\hbar$ is absorbed into the definition of $\alpha_R$ \cite{tanaka2024}. The applied magnetic field induces a homogeneous  Zeeman field $\mathbf{B}$ forming an angle $\theta$ with the $x$-axis, such that $B_x = B \cos\theta$ (the third term) and $B_y = B \sin\theta$ (the fourth term) are   components perpendicular and parallel to the SO axis, respectively; see Fig.~\ref{Fig1}(a). The fifth term represents the proximity induced superconductivity, with $\Delta$ being the induced spin-singlet $s$-wave pair  potential.  The $i$-th Pauli matrices $\sigma_i$ and $\tau_i$, with $i = x, y, z$, operate in the electron’s spin and particle-hole subspaces, respectively, while $\sigma_0$ and $\tau_0$ denote the identity matrices. 
 
To understand the role of the Zeeman field, we diagonalize Eq.~(\ref{eq:ham_BDG}) and  in Fig.~\ref{Fig1}(b) we show the energy versus momentum at distinct values of the angle $\theta$ at fixed Zeeman field amplitude $B=1.2B_{\rm c}$, where $B_{c}=\sqrt{\mu^{2}+\Delta^{2}}$ and its physical meaning is discussed below. The eigenvalues are obtained  by solving the $4th$-order secular equation   $E^4 + a E^2 + b E + c = 0$, which is obtained from the eigenvalue problem. Here, $a = -2[ B_x^2 + B_y^2 + \alpha_{R}^2k^2 + \Delta^2 + \xi_k^2 ]$, $b = -8 B_y \alpha_{R} k \xi_k$, and $c = \xi_k^2[ \xi_k^2 - 2(B^2 + \alpha_{R}^2 k^2 -\Delta^2)] + [ \alpha_{R}^2 k^2 - B_y^2 +\Delta^2]^2 + B^2 [ B_x^2 + 2(B_y^2 + \alpha_{R}^2k^2 - \Delta^2)]$, with  $\xi_k =  (\hbar^2 k^2 )/(2m^*) - \mu$; see Ref.\,\cite{cayao_thesis2017}.  At $\theta=0$, there only exists a  Zeeman field component perpendicular to the SO axis;  the energy versus momentum in this case exhibits two distinct gaps at low momentum and at the Fermi momenta $\pm k_{\rm F}$, here denoted as $\Delta_{1}$ and $\Delta_{2}$ for zero momentum and $+k_{\rm F}$, respectively, see cyan curve in Fig.~\ref{Fig1}(b). As $\theta$ takes finite values, there appears  a finite component $B_{y}$ parallel to the SO axis; the energy versus momentum gets tilted, maintaining constant $\Delta_{1}$ but decreasing $\Delta_{2}$ at $+k_{\rm F}$, see red and green curves in Fig.~\ref{Fig1}(b). When $\Delta=B_{y}$, the outer gap vanishes ($\Delta_{2}=0$) at a critical angle determined by   $\theta_{c}={\rm arcsin}(\Delta/B)$ \cite{cayao2016hybrid}, see red curve   in Fig.~\ref{Fig1}(b);  for our parameters, the critical angle is  $\theta_c \approx 0.1216\pi$. When varying the magnitude of the Zeeman field $B$, both the inner and outer gaps ($\Delta_{1}$ and $\Delta_{2}$), exhibit a strong dependence.  This can be seen in Fig.~\ref{Fig1}(c,d), where we show $\Delta_{1,2}$ as a function of $B$ and $\theta$.
We restrict $\theta$ to the range $0$ and $\pi/2$ because for $\pi/2<\theta<\pi$ region, $\Delta_{2}$ follows the same evolution but in reverse;  $\Delta_{2}$ returns to its initial values at $n\pi \pm \theta$ for $n\in \mathbb{Z}$. 
In this scenario, the inner gap  $\Delta_{1}$ vanishes at  $B=B_{c}$, a situation that is insensitive to   variations of $\theta$ and corresponds to the topological phase transition separating the trivial    and topological phases for ($B<B_{c}$)  and ($B>B_{c}$), respectively; see white region in Fig.~\ref{Fig1}(c). In the topological phase, topologically protected MBSs emerge at zero energy separated from the quasicontinuum by $\Delta_{2}$  and located at the ends of the system with a localization length   defined by $\sim 1/\Delta_{2}$ \cite{tanaka2024}.   Moreover, as already anticipated in  Fig.~\ref{Fig1}(b), the outer gap $\Delta_{2}$ decreases as $\theta$ takes finite values and vanishes at the critical angle that depends on $B$ at fixed $\Delta$, see white region in Fig.~\ref{Fig1}(d) where the red and blue colors indicate that  the outer gap takes positive ($\Delta_2>0$) and negative ($\Delta_2<0$) values. A positive $\Delta_{2}>0$ implies that the positive band closest to zero  at $k=k_F$ is above zero (the Fermi level), while $\Delta_{2}=0$ and $\Delta_{2}<0$ indicate  that such band touches zero energy and  goes below zero energy, respectively, when $\theta$ changes; see Fig.~\ref{Fig1}(b). The reduction of $\Delta_{2}$ suggests that the topological protection of localization of MBSs is affected by the   Zeeman field parallel to the SO axis. Despite the apparent detrimental effect of the parallel Zeeman field $B_{y}$, we will show below that it plays an important role for realizing the JD effect. In fact, from a symmetry perspective, the Rashba term breaks the inversion symmetry, while the Zeeman terms break the time-reversal symmetry. Notably, the Zeeman field component $B_{y}$  also   breaks the $x$-inverting symmetry and, as we will see below, is responsible for inducing the JD effect \cite{he2024}  provided $B_x$ is present in the system.
 
To study the JD effect in JJs formed by single-channel semiconducting nanowires, we discretize Eq.\,(\ref{eq:ham_BDG}) into a tight-binding lattice with a lattice constant $a$ divided into three regions of finite length, as sketched in Fig.\,\ref{Fig1}(a). The central region (denoted by N) has no superconductivity ($\Delta = 0$), while the left and right regions (denoted by S$_{\rm L/R}$)  have  proximity-induced superconductivity characterized by    $\Delta\,e^{\pm i\phi/2}$; here  $\Delta$ is the spin-singlet $s$-wave pair potential and $\pm\phi/2$ are the superconducting phases that lead to a finite phase difference across the junction and enable the study of the Josephson effect \cite{cayao2018,tanaka2024}. Thus, the Hamiltonian for the JJ can be written as
\begin{equation}\label{eq:H_SNS}
	H_{\rm SNS} = H_{\rm S_L} + H_{\rm N} + H_{\rm S_R} + H_{\rm T}\,,
\end{equation}
where $H_{\rm S_L}$, $H_{\rm N}$, $H_{\rm S_R}$ describe the left superconductor $\mathrm{S_L}$, the normal region N, and the right superconductor $\mathrm{S_R}$, respectively. The last element $H_{\rm T}$ in Eq.\,(\ref{eq:H_SNS}) is the Hamiltonian corresponding to the coupling between the superconductors and normal region. To be more explicit, $H_{\rm S_\alpha}$ and $H_{\rm N}$, and $H_{\rm T}$ are given by   
\begin{eqnarray}
H_{\rm S_{\alpha}} &=& \sum_{n} c_{n_\alpha}^\dagger h_{nn} c_{n_\alpha} + \sum_{\langle n,m \rangle} c_{n_\alpha}^\dagger V_{nm} c_{n_\alpha} \nonumber\\
		&+&\sum_{n} \Delta e^{i\phi_\alpha} (c^{\dagger}_{n_\alpha, \uparrow} c^{\dagger}_{n_\alpha, \downarrow} - c^{\dagger}_{n_\alpha, \downarrow} c^{\dagger}_{n_\alpha, \uparrow}) + \mathrm{h.c.}\,,\label{eq:H_S}\\
		H_{\rm N} &=& \sum_{j} c_{j_\mathrm{N}}^\dagger h_{jj} c_{j_\mathrm{N}} + \sum_{\langle j,k \rangle} c_{j_\mathrm{N}}^\dagger V_{jk} c_{k_\mathrm{N}}+ \mathrm{h.c.},\,,\label{eq:H_N}\\
H_{\rm T} &=& V_{\rm T} (c_{M_\mathrm{L}}^\dagger c_{1_\mathrm{N}} + c_{M_\mathrm{N}}^\dagger c_{1_\mathrm{R}})+ \mathrm{h.c.},\label{eq:H_T}
\end{eqnarray}
where  $n\in\{1_\alpha, 2_\alpha, \dotsc , M_\alpha \}$ labels the lattice sites in  $\mathrm{S}_{\alpha} (\alpha = L, R)$ and $c_{n_\alpha} = (c_{n_\alpha, \uparrow}, c_{n_\alpha, \downarrow})^\mathrm{T}$ annihilates an electronic state at site $n$  with spin $\sigma=\uparrow,\downarrow$ in $\mathrm{S}_{\alpha} (\alpha = L, R)$. Similarly $j\in \{1_\mathrm{N}, 2_\mathrm{N}, \dotsc , M_\mathrm{N} \}$ denotes  the sites in the N region  and the operator $c_{j_\mathrm{N}} = (c_{j_\mathrm{N}, \uparrow}, c_{j_\mathrm{N}, \downarrow})^\mathrm{T}$ annihilates an electronic state at site $j$  with spin $\sigma=\uparrow,\downarrow$. Thus, the lengths of the N and  ${\rm S_\alpha}$ regions are  $L_{\rm N}=aM_{\rm N}$ and $L_{\rm S}=aM_{\alpha}$.
Moreover,  as anticipated, $\phi_\alpha$ in Eq.\,(\ref{eq:H_S}) is considered to have  constant values of $+\phi/2$ and $-\phi/2$ in ${\rm S_L}$ and ${\rm S_R}$, respectively. Here, the onsite terms  ($h_{nn}$) and hopping ($V_{nm}$) matrices are given by   
 \begin{equation}
 \begin{split}
 h_{nn}  &= (2t - \mu_{n})\sigma_0 + B_x \sigma_x + B_y \sigma_y\,,\\
 V_{nm} &= -t \sigma_0 + i t_{\rm SO} \sigma_y\,,
 \end{split}
 \end{equation}
where $t = \hbar^2/(2m^*a^2)$ is the hopping, $t_{\rm SO} =  \alpha_{\rm R}/(2a)$ is the Rashba SOC hopping, and $\mu_{i}$ is the chemical potential which we keep equal in all the regions. The Zeeman field $\mathbf{B} = (B_x, B_y)$ is applied uniformly throughout the entire junction, penetrating all three regions identically.
Furthermore, since $H_{\rm T}$ couples $\mathrm{S}_{\rm L,R}$ and N regions, it contains finite entries only for the adjacent sites that are located at the interfaces between the $\mathrm{S}_{\rm L,R}$ and N regions; they are  determined by $V_\mathrm{T} = \tau V_{nm}$, where $\tau$ controls the transparency of the junction with $\tau\in[0, 1]$. In this way, $\tau$ tunes $V_{\rm T}$ and thereby the normal transmission $T_{\rm N}$. Specifically, $\tau = 1$ corresponds to a fully transparent junction ($T_{\rm N} = 1$), while the tunneling regime ($T_{\rm N} \ll 1$) is reached for sufficiently small $\tau$. To treat electrons and holes at the same level, we write Eq.\,(\ref{eq:H_SNS}) in Nambu space with the Nambu spinor   $\Psi = (c_\mathrm{S_L}, c_{\rm N}, c_{\rm S_R}, c_\mathrm{S_L}^\dagger, c_{\rm N}^\dagger, c_{\rm S_R}^\dagger)^\mathrm{T}$, where $c_\mathrm{S_\alpha}$ and $c_{\mathrm{N}}$ span   all the lattice sites in   $\mathrm{S_\alpha}$, with operators $c_{n_\alpha}$,  and in N, with operators $c_{j_\mathrm{N}}$, respectively.   Hence, the JJ Hamiltonian Eq.\,(\ref{eq:H_SNS}) in Nambu space is given by
   \begin{equation}\label{eq:H_BdG_discretized}
		   	{\cal H}_{\rm SNS} = \frac{1}{2}\Psi^\dagger H_{\rm BdG} \Psi\,,
		   \end{equation}
where $H_{\rm BdG}$ is the BdG Hamiltonian in real space which we later diagonalize to explore the energy spectra and supercurrents.

In our simulations,  we use $a=10$\,nm and consider realistic parameters such as the electron's effective mass $m^* = 0.015 m_e$, the strength of Rashba SOC   $\alpha_R = 20$\,meV-nm, and superconducting gap $\Delta = 0.25$\,meV, within the range of experimental values given for InSb or InAs semiconductors and Al or Nb superconductors, see e. g., Ref.\,\cite{lutchyn2018}. Additionally, we fix the chemical potential  at $\mu=0.5$\,meV. We also consider Zeeman fields up to $2B_c$, whose associated magnetic fields $\mathcal{B}$ follow from   $B=g\mu_{\rm B}\mathcal{B}/2$, where $g$ is the Land\'{e} $g$-factor, and $\mu_{\rm B}$ is the Bohr Magneton. Thus, for large $g$-factors, such as for InSb ($g=40$), one obtains magnetic fields $\mathcal{B}$ below the critical field of a conventional superconductor \cite{lutchyn2018}. This is exactly the reason why experiments involving Majorana devices always consider semiconductors with large $g$-factors, see Refs.~\cite{lutchyn2018, prada2020}. Taking into account these realistic parameters, we next model JJs and explore  the formation of ABSs and MBSs in the low energy spectrum and identify under which conditions a JD effect emerges.
 
\begin{figure}[!t]
\centering
 \includegraphics[width=0.99\columnwidth]{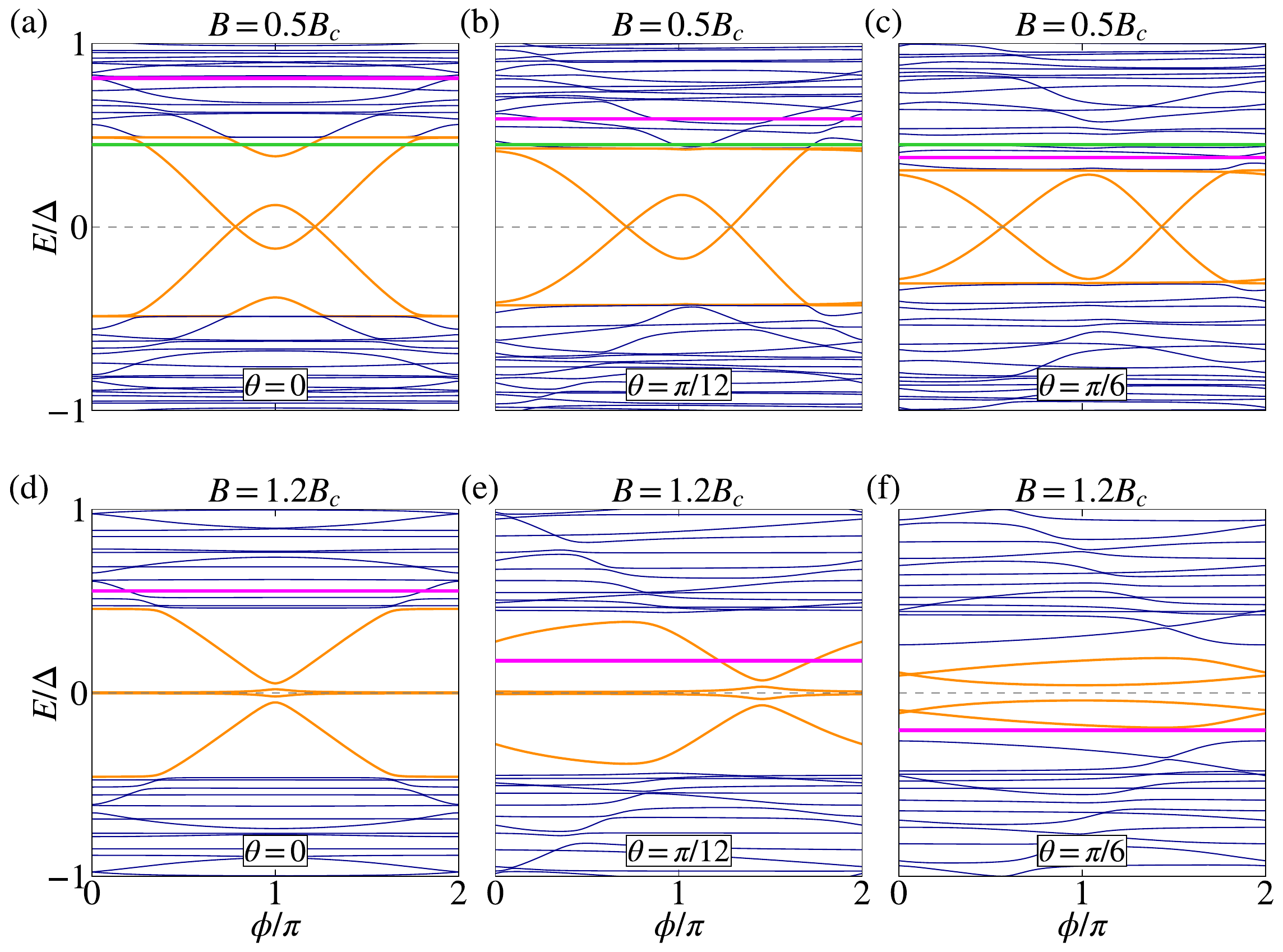}
 \caption{(a-f) Low-energy spectrum as a function of the superconducting phase difference $\phi$ in the trivial (a-c) and topological phases (d-f) at distinct values of $\theta$. The four energy levels  closest to zero energy $E=0$ are indicated in  orange color. The green and magenta horizontal lines represent the inner ($\Delta_1$) and outer ($\Delta_2$) gaps, respectively, obtained from the bulk model given by Eq.\,(\ref{eq:ham_BDG}). In the bottom panels $\Delta_{1}$, appears at higher energies, that is, $\Delta_{1}>\Delta$. $\Delta_{2}$ in (f) lies below $E=0$. Parameters: $L_{\rm S} = 2\,\mu$m and $L_{\rm N}=20$\,nm, $\tau=1$, while the rest of parameters are the same as in Fig.\,\ref{Fig1}.}
\label{Fig2} 
\end{figure}

\section{Low-energy spectrum: emergence of Andreev and Majorana states}
\label{sec:abs}
We start by analyzing the low-energy spectrum of the JJ modelled by Eq.~(\ref{eq:H_BdG_discretized}), taking into account a short junction with $L_{\rm N} = 20$\,nm and $L_{\rm S} = 2\,\mu$m; for long N regions, see App.\,\ref{app:JD_with_long_N}. The motivation to choose   short junctions  relies on that, in this regime, only a few ingap states appear within the induced superconducting gap, as we see next. In Fig.\,\ref{Fig2} we present the low-energy spectrum as a function of the superconducting phase difference $\phi$ in the trivial and topological phases for distinct $\theta$; the energy levels within the induced superconducting gap are indicated by  orange color, while the bulk gaps  $\Delta_{1(2)}$   discussed in previous section are marked by green (magenta) horizontal lines. Since ${\rm min}(\Delta_{1},\Delta_{2})$ naturally defines the lowest gap in the bulk [Fig.\,\ref{Fig1}], all the levels below such a minimum gap are referred to as ingap levels; the explicit behavior of these ingap states strongly depends on the direction of the Zeeman field $\theta$; see see the magenta and green curves in Fig.\,\ref{Fig2}.  Moreover, we note that the dense set of levels above the minimum gap corresponds to the quasicontinuum, which appears because of the finite size systems.

 The first observation in Fig.\,\ref{Fig2} is that the spectrum develops a strong dependence on $\phi$ and exhibits several interesting features depending on the value of $\theta$. When $\theta = 0$, the Zeeman field is oriented perpendicular to the SO axis and the phase-dependent energy spectrum exhibits a symmetric profile with respect to $\phi=\pi$. In this case, the  spectrum in the trivial phase ($B<B_{c}$) hosts one pair of ABSs at  finite $B$ which develops zero energy crossings around $\phi=\pi$ as a result of SOC, see Fig.\,\ref{Fig2}(a) and also Refs.\,\cite{beenakker1992, sauls2018}.  In the topological phase ($B>B_{c}$) shown in Fig.\,\ref{Fig2}(d), the ingap ABSs are topological and define the formation of four MBSs: At $\phi=0$, the lowest almost dispersionless energy levels define a pair of MBSs located at the outer ends of the left and right superconductors, while at $\phi=\pi$ the system hosts two additional MBSs at the inner sides of the junction, making a total of four MBSs \cite{sanjose2012,cayao2017,cayao2018}. Due to the finite length of the superconductors, the four MBSs develop a finite energy splitting at $\phi=\pi$,  which becomes zero when the length of the superconductors is  larger than twice the Majorana localization length \cite{cayao2017}; see also App.\,\ref{app:JC_with_long_S} for more details. The dependence of this energy splitting  is thus a direct result of MBSs being located at the system's ends  and, hence, a property tied to the inherent Majorana nonlocality \cite{cayao2021,baldo2023,PhysRevB.110.224510,ahmed2024oddfreABS}; see App.\,\ref{app:Localization_MBS_in_JD}. The trivial phase does not exhibit a dependence on the length of the superconductors because it does not host nonlocal quasiparticles \cite{cayao2021,baldo2023,awoga2019,ahmed2024oddfreABS}.

When $\theta \neq 0$,  the Zeeman field acquires a nonzero component parallel to the SO axis $B_{y}$, which  then gives rise to a phase-dependent   low-energy spectrum that is asymmetric with respect to $\phi=\pi$ in the trivial and topological phases, see Fig.\,\ref{Fig2}(b,c,e,f).  This asymmetry becomes more pronounced as $\theta$ (or equivalently $B_y$) increases, impacting both the in-gap energy levels and the quasicontinuum above the induced superconducting gap. In the trivial phase ($B<B_{c}$),  the  asymmetry  in the ABSs is weak, but, as $\theta$ increases,   the induced superconducting gap $\Delta_{2}$ (magenta line) reduces   and the zero-energy crossings happen further away from $\phi=\pi$, see Fig.\,\ref{Fig2}(b,c). Interestingly,  in the topological phase ($B>B_{c}$), MBSs exhibit a much stronger asymmetry when  $\theta\neq0$ in comparison to the trivial regime, see Fig.\,\ref{Fig2}(e,f). A notable effect of $\theta\neq0$ is that the zero-energy splitting of MBSs can occur at phase values other than $\phi = \pi$, as shown in Fig.~\ref{Fig2}(d,e,f), thus unveiling the crucial role of the  Zeeman field parallel to the SO axis. Importantly, the value of $\phi$ where the zero-energy splitting occurs, does not change with the increase in superconductor's length; see App.\,\ref{app:JC_with_long_S}. Another effect of $\theta\neq0$ in the topological phase, already seen in Fig.\,\ref{Fig1}(b) of the previous section,  is that the gap $\Delta_{2}$ reduces and can  even become negative, see  magenta line in Fig.\,\ref{Fig2}(e,f). This implies that   levels from the quasicontinuum lower their energies and can coexist with MBSs [Fig.\,\ref{Fig2}(e,f)], thus damaging the topological protection of MBSs provided by $\Delta_{2}$. Yet another impact of $\theta\neq0$ is that even the zero-energy splitting at  phases other than $\phi = \pi$ are not apparent anymore, with the outer MBS even acquiring finite energies for all phases, see Fig.\,\ref{Fig2}(f) when $\Delta_{2}$ is negative.  Despite all the noted effects of $\theta\neq0$, and hence of the Zeeman field component parallel to the SO axis, we will see later that  the asymmetry in the phase-dependent spectrum  with respect to $\phi=\pi$ is perhaps the most important as it  gives rise to the JD effect.

\begin{figure}[!t]
\centering
 \includegraphics[width=0.99\columnwidth]{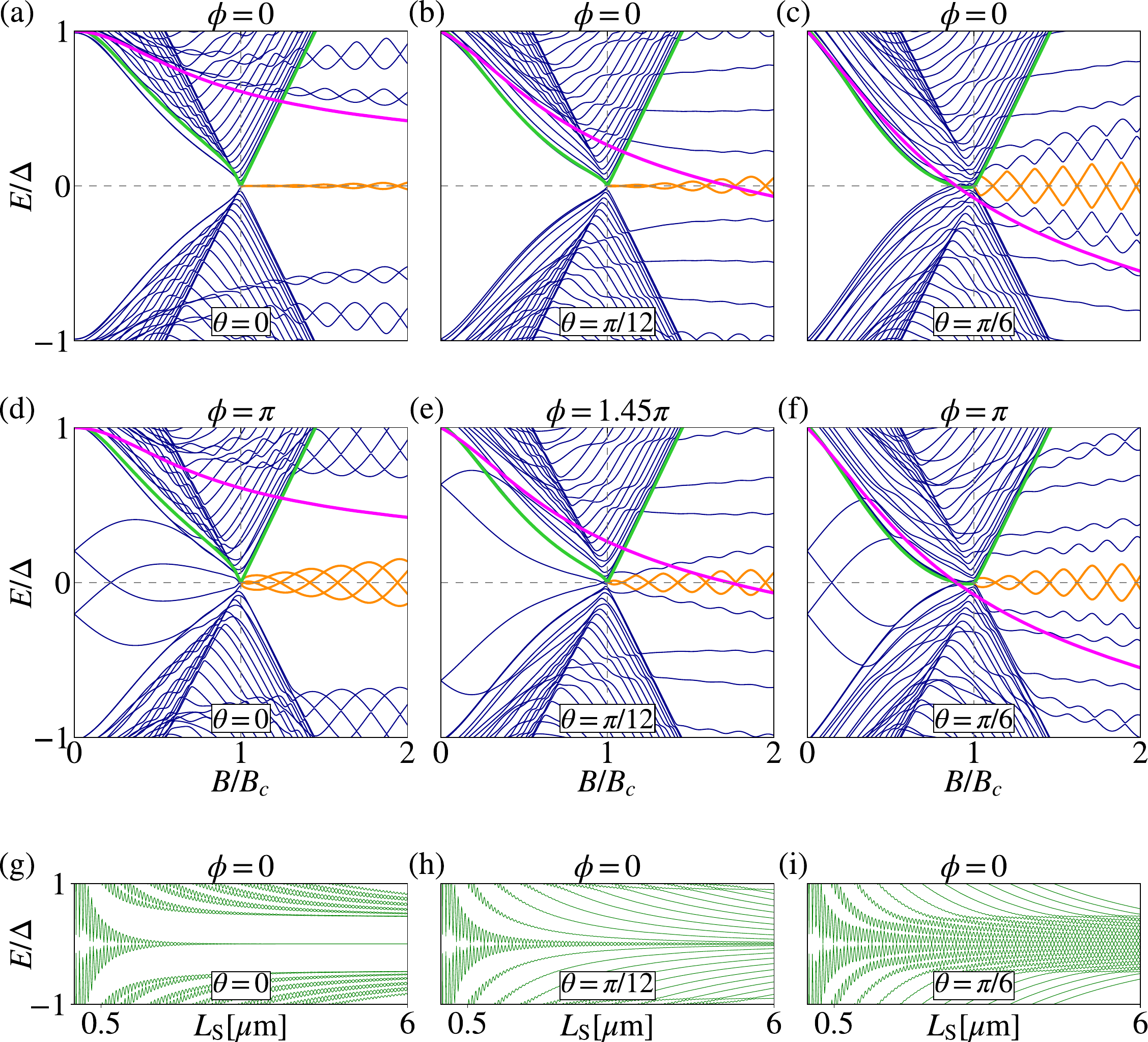}
 \caption{(a-f) Low-energy spectrum as a function of the magnitude of the Zeeman field B for distinct values of $\theta$ at $\phi=0$ (a-c) and  $\phi \neq 0$  (d-f). The choice of  $\phi \neq 0$ in (d-f) corresponds to $\phi$ where the four MBSs split in energy, see  Fig.\,\ref{Fig2}(d-f). In (f), $\phi = \pi$ is selected since the exact splitting point of the MBSs is unclear, see Fig.\,\ref{Fig2}(f). The green and magenta curves show the Zeeman dependence of $\Delta_1$ and $\Delta_2$, respectively. (g-i) illustrate the dependence of the low-energy spectra on $L_\mathrm{S}$, with $B=1.8B_c$. Parameters: $L_{\rm S} = 2\,\mu$m and $L_{\rm N}=20$\,nm, $\tau=1$, while the rest of parameters are the same as in Fig.\,\ref{Fig1}.} 
\label{Fig3} 
\end{figure}

Having understood the phase-dependent energy spectrum, we now analyze the energy spectrum as a function of the Zeeman field, presented in Fig.~\ref{Fig3} at fixed $\phi$ and $\theta$. The values of $\theta$ correspond to the chosen ones in  Fig.~\ref{Fig2}, while the values of   $\phi=0$ and also $\phi$ at which the four MBSs develop the zero-energy splitting.  At $B=0$, the superconducting pair potential produces a gap in the spectrum   irrespective of the value of $\theta$, see Fig.~\ref{Fig3}(a-c). Here,   $\theta$ affects the ingap ABSs appearing at $B = 0$ for $\phi$ other than $\phi=0$, forcing them to even acquire higher (lower) energies, see Fig.~\ref{Fig3}(d-f).  As $B$ takes finite values, the gap in the spectrum reduces and eventually closes at the topological phase transition marked by  $B = B_c$, which is independent of the value of $\theta$ [Fig.~\ref{Fig3}]; the closing of the energy gap closely follows the closing of the bulk gap $\Delta_{1}$, depicted by green curve in Fig.~\ref{Fig3}. As the system transitions into the topological phase $B>B_{c}$ at $\theta=0$, the energy gap reopens and  leaves two (four) energy levels   oscillating around zero energy as $B$ further increases at $\phi=0$ ($\phi=\pi$), see Fig.\,\ref{Fig3}(a,d); these energy levels define the two   MBSs at $\phi=0$ (four   MBSs at $\phi$) discussed in Fig.\,\ref{Fig2}(d) and are within the topological gap $\Delta_{2}$ shown by magenta curve. A finite $\theta$ in the topological phase has a huge impact on the low-energy spectrum, with the most immediate feature being the proliferation of ingap states coming from the quasicontinuum  at energies close to those of MBSs, see  Fig.\,\ref{Fig3}(b,c,e,f). 

Even more dramatically is that the low-energy spectrum in the topological phase can be even gapless at certain values of $\theta$ [Fig.\,\ref{Fig3}(c,f)]; the vanishing of the topological gap becomes more evident in JJs with long superconductors $L_\mathrm{S}$ [Fig.\,\ref{Fig3}(g-i)]. 
For $\theta=0$, there is a pair of energy levels within the lowest gap $\Delta_{2}={\rm min}(\Delta_{1},\Delta_{2})$, although not indicated but it is evident at $E/\Delta\approx0.5$; in fact, the first excited levels develop a clear energy gap separating from the levels closest to zero [Fig.\,\ref{Fig3}(g)]. When $\theta\neq0$, the lowest gap $\Delta_{2}$ gets vanishing values  for the respective Zeeman field $B=1.8B_{c}$, as clearly seen in the magenta curve of  Figs.\,\ref{Fig3}(b,c). That is why there is no clear gap anymore and a gapless spectrum is obtained as $L_\mathrm{S}$ increases [Figs.\,\ref{Fig3}(h,i)],  which is consistent with the vanishing value of $\Delta_{2}$ in Fig.\,\ref{Fig1}(d). Thus the lost of the energy gap in the topological phase is consistent  with the behaviour of the bulk gap at the positive Fermi point $\Delta_{2}$, which, for certain $\theta$, can vanish either in the topological phase [Fig.\,\ref{Fig3}(b,e)] or in the trivial phase [Fig.\,\ref{Fig3}(c,f)]. Thus, Zeeman fields with a finite component parallel to the SO axis via $\theta\neq0$ deteriorate the topological gap that protects MBSs from higher energy states. Despite these seemingly negative effects, we will see in the next sections that $\theta\neq0$ promotes nonreciprocal Josephson transport and hence a JD effect.

 \begin{figure}[!t]
\centering
 \includegraphics[width=0.99\columnwidth]{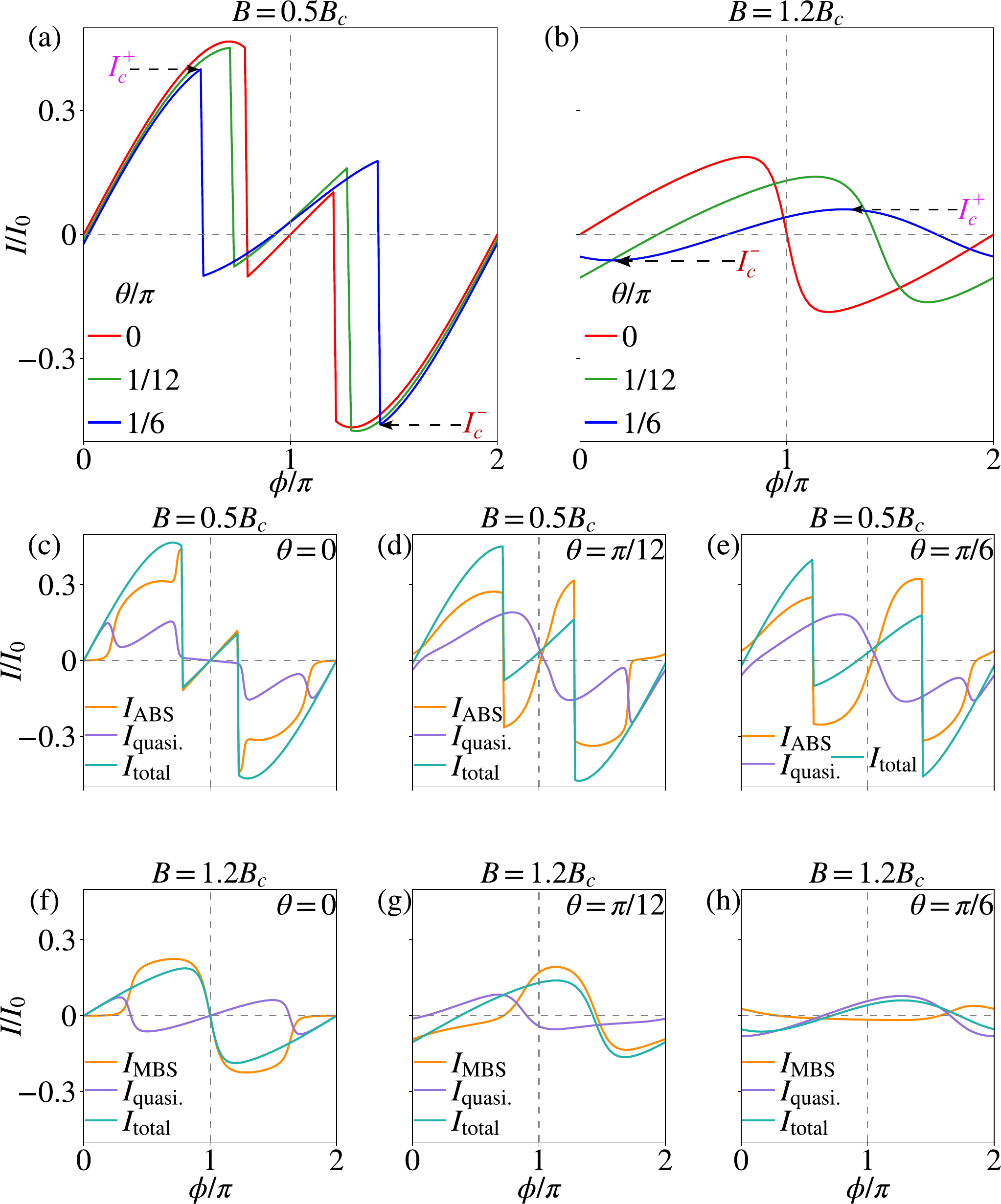}
 \caption{(a,b) Josephson currents as a function of the superconducting phase difference $I(\phi)$ in the trivial (a) and topological (b) phases for distinct $\theta$. The Josephson currents in (a,b) correspond to the phase-dependent spectrum shown in Fig.\,\ref{Fig2}.  (c-e) Contribution of  the ABSs ($I_{\rm ABS}$) and quasicontinuum ($I_{\rm quasi.}$) to the total Josephson current  $I_\mathrm{total}$ corresponding to (a) in the trivial phase. (f-h) Contribution of  the MBSs ($I_{\rm MBS}$) and quasicontinuum ($I_{\rm quasi.}$) to the total Josephson current  $I_\mathrm{total}$ corresponding to (b) in the topological phase. Parameters: $L_{\rm S} = 2\,\mu$m and $L_{\rm N}=20$\,nm, $\tau=1$, $I_0 = e\Delta/\hbar$, $T=0$, while the rest of parameters are the same as in Fig.\,\ref{Fig2}.}
\label{Fig4} 
\end{figure}

\section{Non-reciprocal phase-dependent Josephson currents}
\label{sec:jc}
In this section, we investigate the impact of the Zeeman field oriented at specific angles relative to the SO axis on the  supercurrents across short JJs modeled by Eq.\,(\ref{eq:H_SNS}). At finite temperatures, the Josephson current can be obtained from the phase-dependent discrete energy spectrum as \cite{beenakker1992}
\begin{equation}
I(\phi)  = -\frac{e}{\hbar}\sum_{ \varepsilon_n>0 }^{ }{\rm tanh}\Big[\frac{\varepsilon_{n}}{2\kappa_{\rm B}T}\Big] \frac{d\varepsilon_n(\phi)}{d\phi}\,,
\end{equation}
where $\varepsilon_n(\phi)$ denotes the discrete positive phase-dependent energy levels, $\kappa_{\rm B}$ is the Boltzmann constant, and $T$ the temperature. While we primarily focus on the zero-temperature limit ($T=0$), the results that we present here remain robust at finite temperatures but smaller than $\Delta$, see  App.\,\ref{app:Efficiency_temperature}. We then numerically calculate the Josephson current $I(\phi)$ for distinct $\theta$ across a transparent JJ and present it in Fig.\,\ref{Fig4}  in the trivial and topological phases. In the trivial phase at $\theta=0$, the supercurrent $I(\phi)$ has a regular behavior, where $I(\phi)=-I(-\phi)$ and $I(\phi)=0$ at $\phi=m\pi$, with $m\in\mathbb{Z}$, and developing a sawtooth-like profile at phases around $\phi=\pi$ coming from the SOC effect in the spectrum, see red curve in Fig.\,\ref{Fig4}(a).   With the introduction of $\theta \neq 0$, and hence of $B_{y}$ parallel to the SO axis, the Josephson current $I(\phi)$ becomes asymmetric with respect to $\phi = \pi$ [Fig.\,\ref{Fig4}(a)], which is a consequence of the asymmetric phase dependent Andreev spectrum shown in Fig.\,\ref{Fig2}(a-c). Another consequence is that, at finite $\theta$, the supercurrent  develops distinct global maximum and minimum $I_{\rm c}^{\pm}={\rm max}_{\phi}[\pm I(\phi)]$, leading to $I(\phi)\neq-I(-\phi)$ as with $\theta=0$. We note that $I_{c}^{\pm}$ are known as critical currents. The profile of $I(\phi)$ is largely determined by the ingap ABSs, although a finite contribution exist due to the quasicontinuum, see Fig.\,\ref{Fig4}(c,d,e).  In the topological regime,  the effect of $\theta\neq0$ remains in the Josephson current, inducing an asymmetric profile with respect to $\phi=\pi$ due to the phase-dependent energy spectrum and different values of the global maximum and minimum $I_{\rm c}^{\pm}$, see Fig.\,\ref{Fig4}(b). Moreover, $I_{\rm c}^{\pm}$ increases with longer superconducting regions \cite{cayao2017}; see App.\,\ref{app:JC_with_long_S}.  The regular properties of $I(\phi)$ with MBSs at $\theta=0$, including the sine-like profile with $\phi$ and $I(\phi)=0$ for $\phi=m\pi$,  are considerably affected in the topological phase when $\theta\neq0$, leading to a $\phi_{0}$-junction behavior that is more pronounced than in the trivial phase. The global maximum and minimum $I_{c}^{\pm}$ in the topological phase are in general smaller than in the trivial phase but  are more susceptible to changes in $B$ and $\theta$ [Fig.\,\ref{Fig4}(a,b)]; hence,   the   values of $\phi$ at which $I_c^\pm$ occur in the topological phase, denoted as $\phi_\pm$, vary more with both $B$ and $\theta$. For instance, for $B = 1.2B_c$, $\phi_+$ shifts from $\phi < \pi$ at $\theta = 0$ to $\phi > \pi$ at $\theta = \pi/12$.  This shift in $\phi_\pm$ correlates with the zero-energy states found in   Fig.\,\ref{Fig2} and Fig.\,\ref{Fig3}. Moreover,  we note that in the topological phase,   the contribution to   $I(\phi)$ is largely dominated by the MBSs even for  $\theta\neq0$, but when the topological gap is vanishingly small, the quasicontinuum develops a  considerable contribution,  Fig.\,\ref{Fig4}(f,g,h).  

Before going further, we highlight that, although $\theta\neq0$ has multiple consequences on $I(\phi)$, the behaviour of $I(\phi)$  exhibiting $I_{c}^{+}\neq I_{c}^{-}$ at $\theta\neq0$ implies that there emerges a non-reciprocal Josephson transport across the JJs considered here. Notably, this nonreciprocal transport behavior signals the emergence of the JD effect, entirely due to $\theta \neq0$ which corresponds to the presence of a Zeeman field component  parallel to the SOC. Thus, the considered JJs exhibit a JD effect in the trivial and topological phases with ABSs and MBSs. Although the consequences of $\theta$ seem to be similar in the trivial and  topological phases, the presence of MBSs in the topological phase makes the JDs susceptible to  their  properties such as the inherent Majorana nonlocality, which we address in the next section.

 \begin{figure}[!t]
\centering
 \includegraphics[width=0.99\columnwidth]{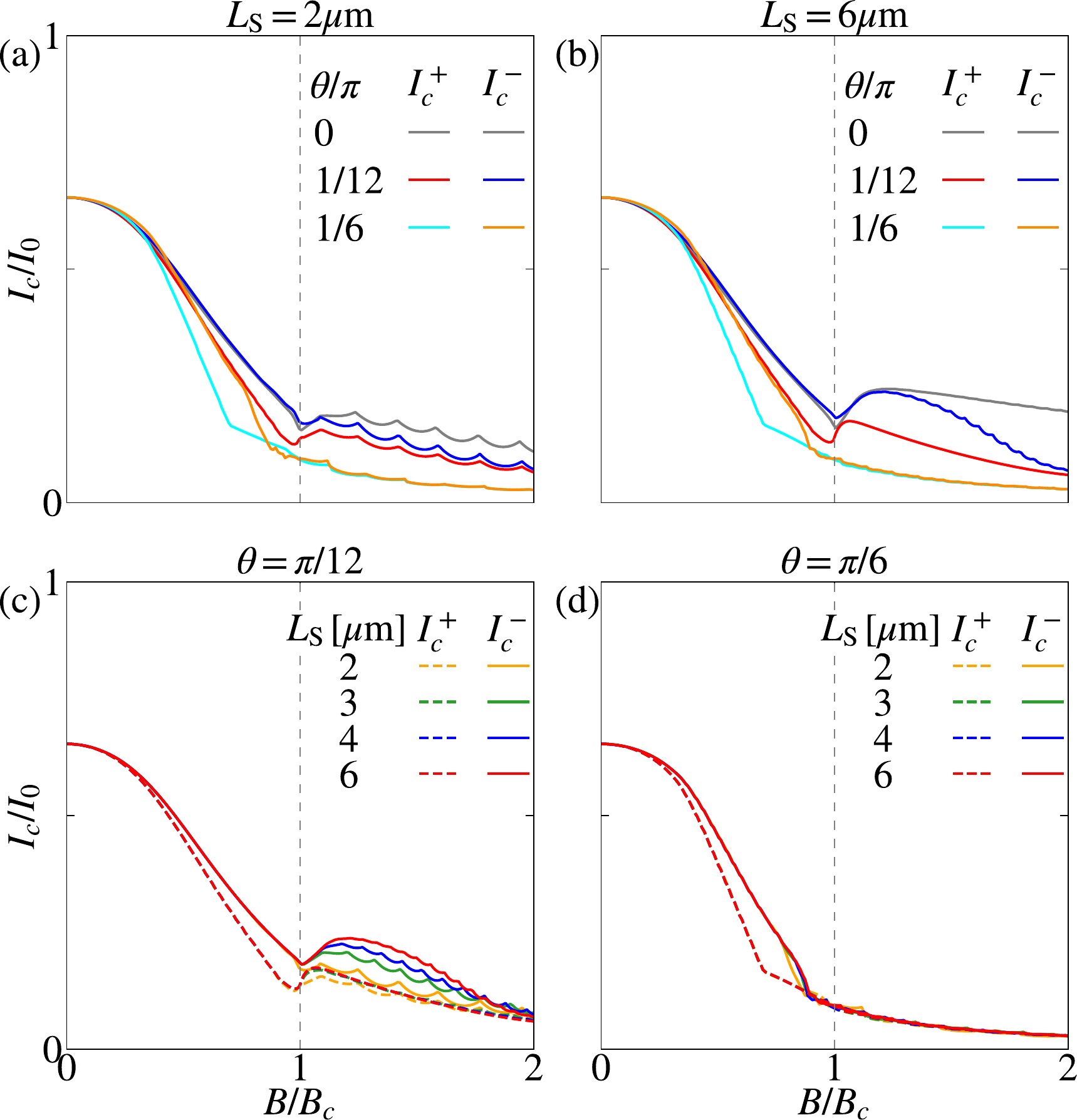}
 \caption{(a,b) Critical currents  $I_{c}^{\pm}$  as a function of the Zeeman field amplitude $B$ for distinct values of $\theta$, and at $L_{\rm S}=2$\,$\mu$m  and  $L_{\rm S}=6$\,$\mu$m. (c,d) The same   as in (a,b) but at fixed $\theta=\pi/12$ and $\theta=\pi/6$ for different $L_{\rm S}$. Parameters:   $L_{\rm N}=20$\,nm, $\tau=1$, $I_0 = e\Delta/\hbar$, while the rest of parameters are the same as in Fig.\,\ref{Fig2}. }
\label{Fig5} 
\end{figure}

\section{Critical currents and Josephson diode's efficiencies}
\label{sec:efficiency}
To further understand the non-reciprocal Josephson transport and characterize the JD effect, here we examine the critical currents $I_{c}^{\pm}$ associated to $I(\phi)$ and also inspect the degree of non-reciprocity that such critical currents exhibit. As already mentioned before, non-reciprocal critical currents signal the emergence of the JD effect.  Below we address the critical currents and diode efficiencies in   JJs studied in the previous section  unless otherwise stated.

\subsection{Non-reciprocal critical currents}
We begin by inspecting the critical currents across transparent short JJs as a function of the Zeeman field amplitude, which is presented in Fig.\,\ref{Fig5}(a,b) for short and long superconducting regions at distinct values of $\theta$. Fig.\,\ref{Fig5}(c,d) shows the Zeeman dependent critical currents at two fixed  $\theta$ for different $L_{\rm S}$. At $\theta=0$, the critical currents $I_c^+$ and $I_c^-$ coincide since the low-energy phase-dependent spectrum is symmetric and $I(\phi)$
 develops  a regular behavior already reported before, see  gray curves in Fig.\,\ref{Fig5}(a,b).  As $B$ increases within   $B<B_{c}$, the critical currents $I_c^\pm$  at $\theta=0$ reduce and, at $B=B_{c}$ when  $\Delta_{1}=0$, they  exhibit a kink-like feature whose finite value arises due to the phase dependence of the low-energy spectrum. In the topological phase   $B>B_{c}$, the critical currents develop   an oscillatory profile as $B$ increases, which originate due to the zero-energy splitting of the four MBSs at $\phi=\pi$  [Fig.\,\ref{Fig2}(d)]. The oscillations in the Zeeman dependent  critical currents are then washed out  when the superconducting regions are longer than twice the Majorana localization length, see gray curve in Fig.\,\ref{Fig5}(b); this situation   also enhances the critical current since for a superconductor of infinite length, the  critical current is due to $\Delta_{2}$.  Hence, the critical currents trace the gap closing and reopening of $\Delta_{1}$ as well as the emergence of MBSs protected by $\Delta_{2}$.

When $\theta\neq0$, we find that the critical currents can be distinct, namely, $I_{c}^{+}\neq I_{c}^{-}$, thus reflecting non-reciprocal Josephson transport, see Fig.\,\ref{Fig5}(a,b). Despite the effect of $\theta$, both critical currents still exhibit   features of the gap closing and reopening at $B=B_{c}$ and also develop the oscillations of MBSs in the topological phase which reduce for long superconductors. An interesting feature to note, and which might  affect the identification of the topological phase, is that $I_{c}^{+}$ tends to decrease faster with $B$ in the trivial phase for certain $\theta$, originating a large difference between $I_{c}^{+}$  and $I_{c}^{-}$, see e. g., red (cyan) and blue  (orange) curves in  Fig.\,\ref{Fig5}(a). The finite values of $\theta$ also soften the   critical current oscillations due to MBSs [Fig.\,\ref{Fig5}(a)] and   tend to reduce the critical current values in the topological phase, see Fig.\,\ref{Fig5}(a,b). The reduction of the critical current in the topological phase is associated to the decrease (or even vanishing) of the topological gap separating MBSs from the quasicontinuum, which becomes more evident when the  superconductor length $L_{\rm S}$ is very long  as then the topological gap is given by the bulk gap $\Delta_{2}$ [Fig.\,\ref{Fig5}(b)]. 

Further insights on the length of the superconductors is obtained from Fig.\,\ref{Fig5}(c,d), where the critical currents are shown for distinct $L_{\rm S}$ at fixed $\theta$. For certain $\theta\neq0$ with an already present non-reciprocity, increasing the length $L_{\rm S}$ enhances the difference between  $I_{c}^{+}$  and $I_{c}^{-}$ in the topological phase but leaves unchanged the critical currents in the trivial phase--see Fig.\,\ref{Fig5}(c) for  $\theta = \pi/12$; in Fig.\,\ref{Fig5}(d), the response of the critical currents to variations in $L_\mathrm{S}$ is negligible mainly because MBSs are almost dispersionless with $\phi$ [Fig.\,\ref{Fig2}(f)]. The dependence on the length of the superconductors can only be attributed to the spatial nonlocality of  MBSs since they are located at the ends of the superconductors. Additionally, the period of the critical current oscillations increases with $L_\mathrm{S}$, which is attributed to the growing number of Majorana zero-energy crossings in the topological regime \cite{cayao2017}. The non-reciprocity in the critical currents can thus sense the emergence of MBSs, while MBSs are able to enhance the diode behavior. Taking into account the above discussion, the found non-reciprocity in the critical currents $I_{c}^{\pm}$ at $\theta\neq0$ demonstrates the emergence of the JD effect with ABSs and MBSs, and its controllability  by the amplitude of the Zeeman field $B$.

 \begin{figure*}[!t]
\centering
 \includegraphics[width=\linewidth]{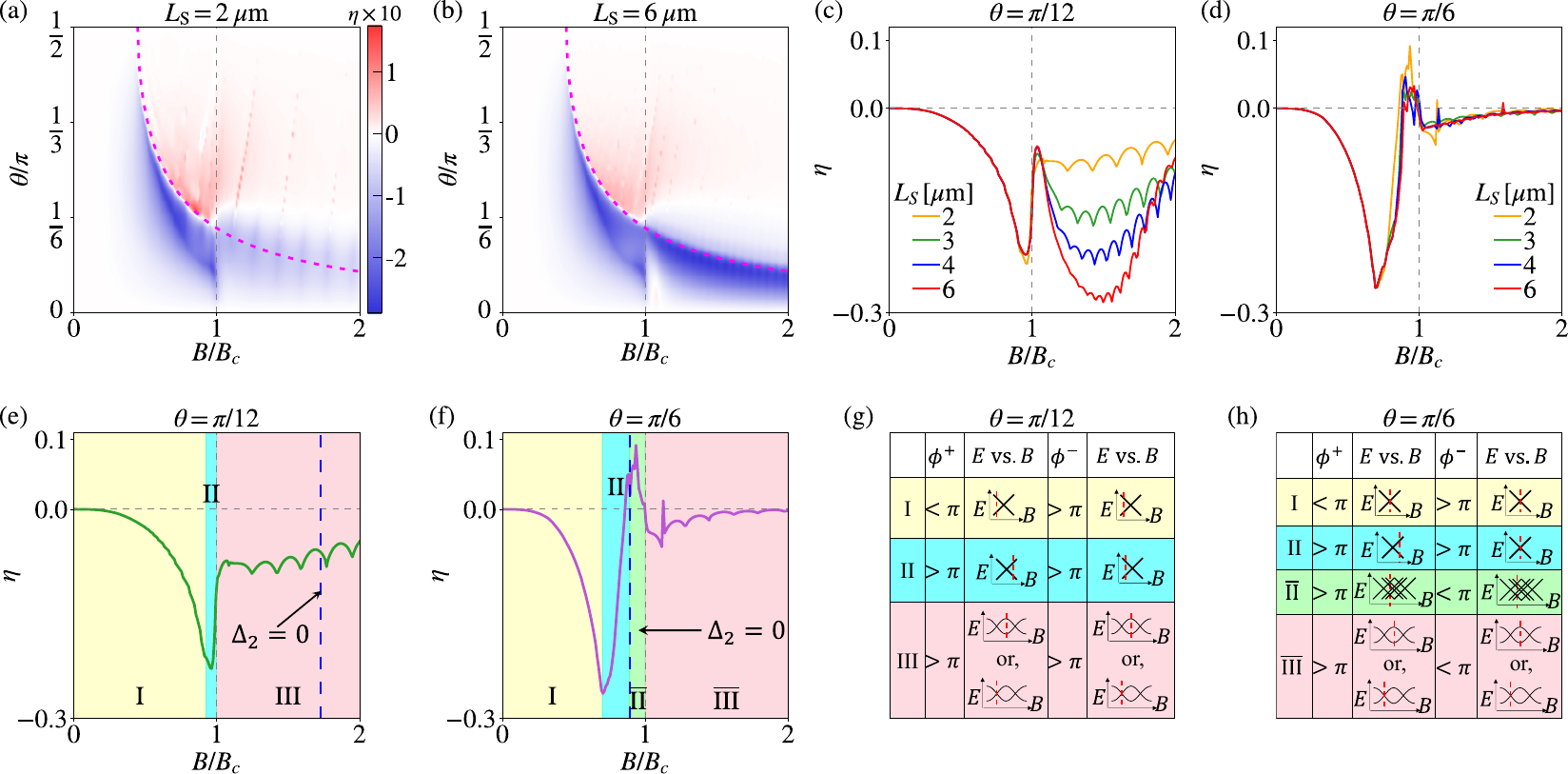}
 \caption{(a,b) Quality factor $\eta$ of the JDs as a function of the Zeeman field amplitude $B$ and $\theta$ in JJs with short ($L_{\rm S}=2$\,$\mu$m) and long ($L_{\rm S}=4$\,$\mu$m) superconductors. The vertical dashed gray line marks the topological phase transition $B=B_{c}$ where the bulk gap $\Delta_{1}$ vanishes, while the dotted magenta curve corresponds to the vanishing of the bulk gap $\Delta_{2}$. (c,d)  Quality factor $\eta$ versus $B$  at two values of $\theta$ in (a,b), including their evolution as the length of the superconductors $L_{\rm S}$ gets longer.  (e,f)   $\eta$ versus $B$ from (a) at $\theta=\pi/12$ and $\theta=\pi/6$ for  $L_{\rm S}=2$\,$\mu$m, indicating distinct regions  that correspond  to the behavior of $\eta$. The vertical  dashed blue lines indicate the points where the bulk gap $\Delta_{2}=0$. (g,h) Table showing the phases $\phi_{\pm}$ at which $I_{c}^{\pm}$  occur in each region of (e,f), accompanied with a sketch of their respective energy versus $B$ around zero energy. In each $E$ vs. $B$ of (g,h), the red dashed line indicates the value of $B$ at which $\eta$ is evaluated. Parameters:   $L_{\rm N}=20$\,nm, $\tau=1$, while the rest of parameters are the same as in Fig.\,\ref{Fig2}.}
\label{Fig6} 
\end{figure*}

\subsection{Efficiency of the Josephson diode effect}
To further understand the emergence of the JDs, in this section we characterize the amount of non-reciprocity of the critical currents  by the quality factor $\eta = (I_c^+ - I_c^-)/ (I_c^+ + I_c^-)$.  
Here, the parameter $\eta$ quantifies the directional asymmetry of the supercurrent across the junction. A positive value of $\eta$ implies that the magnitude of the critical current is larger when flowing from the left to the right superconductor, whereas a negative $\eta$ indicates that the dominant supercurrent flows in the opposite direction—from right to left. In this context, $\eta$ serves as a measure of the (nonreciprocal) diode  behavior of the junction, where the sign indicates the preferred current direction. Moreover, a small (large) value of $\eta$ indicates that one transport direction is preferred over the other, thus making it possible to quantify the amount of nonreciprocity that tells how good is the diode. That is why $\eta$ is often called diode's efficiency.

In Fig.\,\ref{Fig6}(a,b) we present the quality factor $\eta$ as a function of the Zeeman field amplitude $B$ and $\theta$ in transparent single-channel short JJs with short and long superconductors. For JJs with longer N regions, see App.\,\ref{app:JD_with_long_N}.  The first feature we notice  is that $\eta$ acquires finite values when both $B$ and $\theta$ are nonzero, reflecting the key role of the Zeeman field $B_{y}$ parallel to the SO axis, see blue and red regions in Fig.\,\ref{Fig6}(a,b).  As $B$ increases, the diode's efficiency first acquires nonzero values $\eta\neq0$  in the trivial phase and, as $B$ drives the system into the topological phase, $\eta$   remains finite but with a profile that is tied to the presence of MBSs. In fact, in the topological phase,  $\eta$ as a function of $B$ has    an oscillatory profile that depends on $L_{\rm S}$ and reveals the presence of MBSs, see Fig.\,\ref{Fig6}(a,b). An interesting consequence of MBSs in the topological phase  is that  when MBSs become truly zero modes, e. g., in JJs with very long superconductors, the diode's efficiencies get enhanced entirely due to the spatial Majorana nonlocality; in the trivial phase,   $\eta$  is not altered by $L_{\rm S}$ since there are no nonlocal quasiparticles. The enhancement of $\eta$ by means of Majorana nonlocality is also seen in Fig.\,\ref{Fig6}(c), where we show $\eta$ versus $B$ for distinct values of $L_{\rm S}$ at $\theta=\pi/12$; in this regime,   MBSs still disperse with $\phi$  [Fig.\,\ref{Fig2}(e)] and   the bulk gap $\Delta_{2}$ vanishes deep in the topological phase  [Fig.\,\ref{Fig1}(d)]. At $\theta=\pi/6$,   MBSs are   dispersionless almost coexisting with the quasicontinuum [Fig.\,\ref{Fig2}(f)] and no effect of the Majorana nonlocaliy on $\eta$ is observed [Fig.\,\ref{Fig6}(d)]; here the bulk gap $\Delta_{2}$ vanishes before $B_{c}$.  Thus, when $\Delta_2$ vanishes in the topological regime (as in the case of $\theta = \pi/12$), the diode efficiency increases with superconductor length. Conversely, when $\Delta_2$ vanishes in the trivial regime (as in the case of $\theta = \pi/6$), $\eta$ remains unchanged regardless of $L_\mathrm{S}$.

Another feature we highlight in  Fig.\,\ref{Fig6}(a,b) is that $\eta$ exhibits positive and negative values, which, surprisingly, occur in both the trivial and topological phases; hence, the reversal of diode's polarity occurs with ABSs and  MBSs in the trivial and topological phases, respectively.  
Further, for transparent JJs with short superconductors, the maximum value of $\eta$ occurs below $B_{c}$ for all $\theta$, with the position of this maximum shifting away from $B = B_c$ as $\theta$ increases. Notably, positive efficiencies $\eta>0$ in the trivial regime appear only when $\Delta_2$ vanishes in this regime, which we demonstrate by showing that the locus of $\eta = 0$ follows  the curve where $\Delta_2=0$, as indicated by the magenta dotted line in Fig.\,\ref{Fig6}(a,b). The curve where the diode's polarity changes sign in the trivial phase does not depend on $L_{\rm S}$, which unveils that the junction does not host quasiparticles that are nonlocal in space affecting $\eta$ and that  reversing its polarity  is very likely a bulk effect.  In the topological phase, the situation is more intriguing because the change in diode's polarity does not  follow $\Delta_{2}=0$. In short and long superconductors, positive efficiencies appear at low but finite values of $\theta$, while negative efficiencies can happen at much higher $\theta$, in both cases revealing the Majorana oscillations. In contrast to the trivial regime, the efficiencies acquire larger positive values in the topological phase when the superconductors are longer,  an indicator that  MBSs are playing an important role [Fig.\,\ref{Fig6}(b)]. 
Interestingly, the large positive values of $\eta$ in long-superconductor JJs within the topological phase undergo a sudden drop at $\theta$ where $\Delta_2 = 0$, although the signatures of Majorana oscillations persist.
It is worth noting that, while these results hold for single channel JJs, they help anticipate the behavior of $\eta$ in the multichannel case. In particular, inter-subband coupling in multichannel JJs is expected to strongly affect the diode's efficiency, with a very likely enhancement due to the increase in the number of MBSs; see App.\,\ref{app:JD_with_long_N}.

Further insights on the behavior of $\eta$ and its relation to the low-energy spectrum in single channel JJs is shown in Fig.\,\ref{Fig6}(e,f) at $L_{\rm S}=2$\,$\mu$m, where the Zeeman evolution of $\eta$ for $\theta=\pi/12$ and $\theta=\pi/6$ is divided into distinct regions   according to the phases $\phi_{\pm}$ at which the critical currents $I_{c}^{\pm}$ in the current-phase curves $I(\phi)$ occur,   see Fig.\,\ref{Fig6}(g,h). It should be noted that $\phi_+$ and $\phi_-$ evolve continuously with variations in $B$ and $\theta$. Thus, we divide the efficiency curve for $\theta = \pi/6$ and $\theta = \pi/12$ into four and three distinct regions, respectively, see Fig.\,\ref{Fig6}(e,f); remember that for $\theta = \pi/6$ the bulk gap $\Delta_{2}$ vanishes just below $B_{c}$, while for $\theta = \pi/12$ it vanishes deep in the topological phase.  The energy versus $B$  shown in Fig.\,\ref{Fig6}(g,h) is obtained at the phases $\phi_{\pm}$ corresponding to  $I_{c}^{\pm}$. 
In the case of $\theta=\pi/12$, $\phi_+$ and $\phi_-$ emerge before the appearance of zero-energy ABSs, see region I depicted in yellow in Fig.\,\ref{Fig6}(e,g). As $B$ increases, $\phi_+$ shifts beyond $\pi$ and $I_c^+$ appears away from the ABSs but $\phi_-$ remains at the $B$ position of the zero-energy ABS, see region II in Fig.\,\ref{Fig6}(e,g); in the topological phase with MBSs, $\phi_{\pm}$ remains above $\pi$, see region III in Fig.\,\ref{Fig6}(e,g).
For $\theta = \pi/6$ in Fig.\,\ref{Fig6}(f,h), the efficiency curve consists of four regions due to the closing of $\Delta_{2}$ in the trivial regime.
In the trivial phase, $\phi_{\pm}$ in regions I and II closely resembles that observed for $\theta = \pi/12$, but the additional region, $\overline{\mathrm{II}}$ is slightly distinct. In this region, the diode's polarity is reversed. 
Near to the topological phase transition but above $B$ where $\Delta_{2}=0$, $\phi_-$ remains below $\pi$, and multiple zero-energy levels appear in the $E$ vs. $B$ spectrum due to the vanishing of $\Delta_2$ at the onset of this region, see region $\overline{\mathrm{II}}$ in Fig.\,\ref{Fig6}(f,h).  The region corresponding to the topological phase, that is, region $\overline{\mathrm{III}}$ is different from the corresponding region  III at $\theta = \pi/12$, $\phi_{+(-)}$ at $\theta = \pi/6$ are above (below) $\pi$, see Fig.\,\ref{Fig6}(f,h).  Consequently, in transparent JJs, the JD effect due to distinct critical currents ($I_c^\pm$) may arise either in the presence or absence of trivial ABSs or topological MBSs.

\subsection{Majorana-only Josephson diode effect controlled by normal transmission}
\begin{figure}[!t]
\centering
 \includegraphics[width=\linewidth]{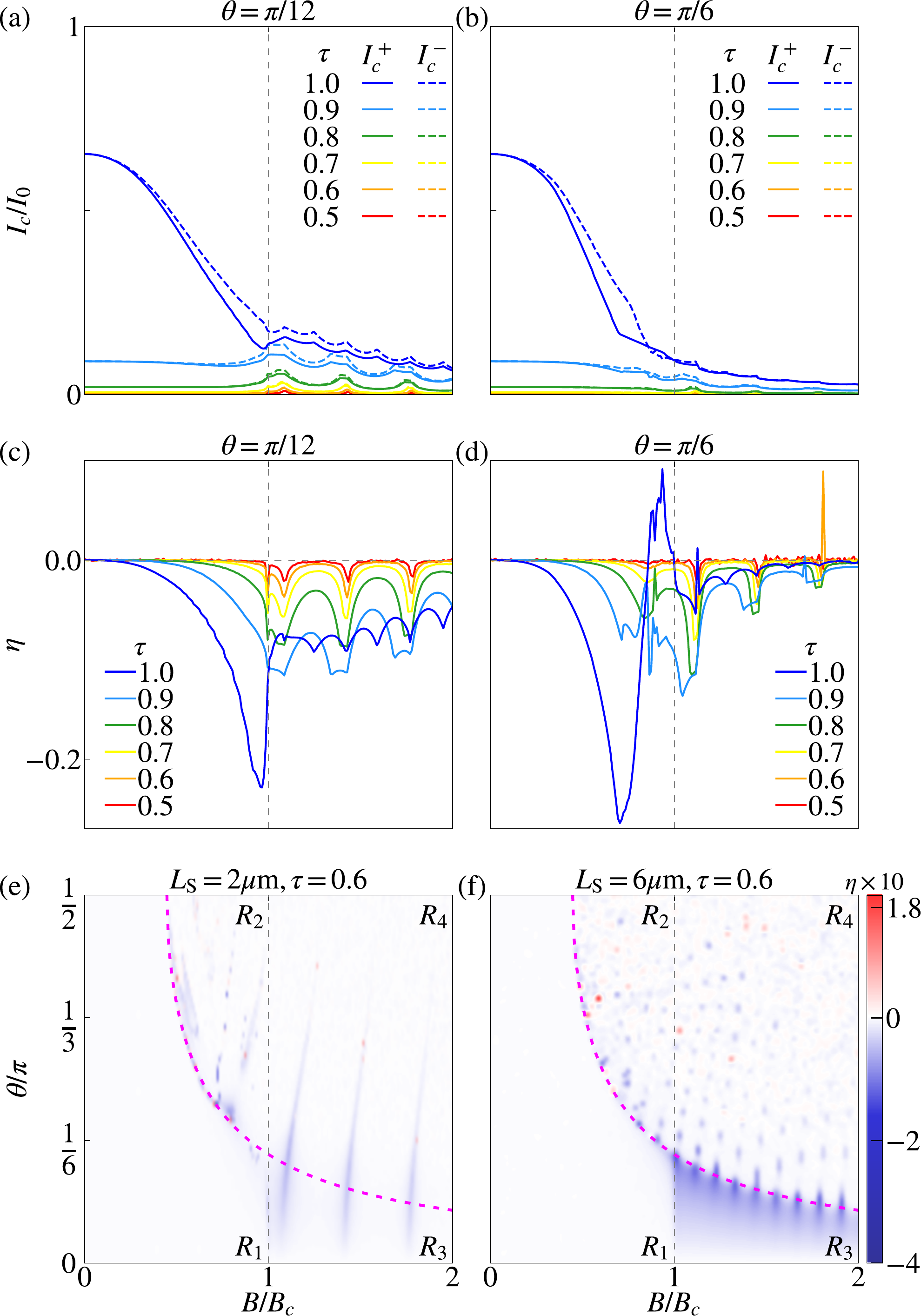}
 \caption{(a,b) Critical currents and quality factors (c,d) as a function of $B$ at finite $\theta$ and different values of $\tau$. The solid and dashed curves in (a,b) correspond to $I_c^+$ and $I_c^-$, respectively, while the vertical dashed gray line marks $B=B_{c}$. (e,f) The quality factor $\eta$ as a function of both $B$ and $\theta$ at $\tau = 0.6$ in JJs with short and long superconductors. The magenta dotted curve represents the vanishing of $\Delta_2$, while the gray dashed line signifies the vanishing of $\Delta_1$.   Here, $R_{1,3}$ ($R_{2,4}$)  regions have $\Delta_2>0$ ($\Delta_2<0$); also, $R_{1,2}$  and $R_{3,4}$ are in the trivial ($B<B_{c}$) and topological ($B>B_{c}$) regimes, respectively. Parameters: the same as in Fig.\,\ref{Fig6}.}
\label{Fig7} 
\end{figure}

All of the preceding results, showing the emergence of JDs in the trivial and topological phases with ABSs and MBSs, respectively, pertain to a transparent JJ based on the model given by Eq.\,(\ref{eq:H_SNS}). Motivated by the fact that critical currents due to ABSs and MBSs have a distinct dependence on the normal transmission $T_{\rm N}$, here we explore how the JD effect responds to variations of such a transmission as $B$ transitions from the trivial to the topological phases. As noted in Sec.~\ref{sec:ham}, the parameter $\tau$ between nearest-neighbor sites in the superconducting and normal regions controls the normal transmission:  by reducing $\tau$, the  tunneling  regime $T_{\rm N}\ll1$ is achieved at $\tau\approx0.5$ \cite{cayao2017}, albeit, at $\tau=0.6$, the junction already exhibits $T_{\rm N}\ll 1$; in the transparent regime, $\tau=1$ corresponds to $T_{\rm N}=1$. That being said, in Fig.\,\ref{Fig7}(a-d) we present the critical currents $I_{c}^{\pm}$ and quality factor $\eta$ as a function of $B$ for two representative values of $\theta$ at $L_{\rm S}=2$\,$\mu$m. In Fig.\,\ref{Fig7}(e,f), we show $\eta$ as a function of   $B$ and $\theta$ for two distinct values of $L_{\rm S}$ in the tunneling regime. As   $\tau$ decreases in the trivial phase $B<B_{c}$, the critical currents $I_c^\pm$ get reduced until they become negligible in the tunneling regime at $\tau\approx0.5$.  The kink-like feature  near $B = B_c$  found for transparent JJs [Fig.\,\ref{Fig5}] changes into a step-like profile as $\tau$ reduces, producing a sudden reentrant critical currents at the topological phase transition in the tunneling regime even when such critical currents vanish below $B_{c}$, see Fig.\,\ref{Fig7}(a) for $\theta=\pi/12$.  Interestingly, this re-entrant effect arises due to the distinct contribution to the critical currents coming  either from ABSs in the trivial phase or    MBSs in the topological phase. In fact, the critical current mediated by MBSs at $\theta=0$ scales as $\sim \sqrt{T_{\rm N}}$, while the critical currents due to ABSs are proportional to $\sim T_{\rm N}$, leading to vanishing critical currents in the trivial phase in the tunneling regime. We have verified that this critical current behavior persists at small but nonzero values of $\tau$, thus explaining the vanishing of critical current due to ABSs in the tunneling regime in Fig.\,\ref{Fig7}(a,b). 

Furthermore, the oscillations of the Majorana-driven critical currents for $B>B_{c}$ reduce their periodicity as $\tau$ is reduced, with a doubled   periodicity in the tunneling regime in comparison to the transparent regime, see Fig.\,\ref{Fig7}(a,b). This period doubling is a direct result of the four MBSs  being decoupled in the tunneling regime, such that two MBSs in each superconductor still oscillate with the same period when $B$ increases. A similar behavior is observed for $\theta = \pi/6$, albeit the oscillatory $I_c^\pm$ in the $B > B_c$ region is weaker due to the almost dispersionless energy spectrum with $\phi$ [Fig.\,\ref{Fig2}(f)]. Therefore, in the tunnelling regime, $I_c^\pm$ diminishes everywhere except in the topological phase where MBSs appear, originating Majorana-driven non-reciprocal critical currents and hence a Majorana-only JD effect. 

 When it comes to the efficiency of the   non-reciprocal critical currents in the tunneling regime, $\eta$ exhibits  vanishing values in the trivial phase for $\tau\ll 1$ for any $\theta$, see  Fig.\,\ref{Fig7}(c,d). In the $B > B_c$ regime, $\eta$ shows the oscillatory behavior with doubled periodicity in the tunneling regime irrespective of the value of $\theta$. It is fair to say that, despite having a Majorana-only JD effect,  its resulting efficiency is low due to a decreased difference between $I_c^+$ and $I_c^-$ in the tunneling regime. Further insights on $\eta$ are obtained from Fig.\,\ref{Fig7}(e,f), where we show $\eta$ as a function of both $B$ and $\theta$ in the tunneling regime and for short and long superconductors. In this case, the diode's efficiency is found to vanish $\eta=0$ in the trivial phase $B<B_{c}$ only when $\Delta_2 > 0$, see region $R_{1}$ below the magenta dotted curve in  Fig.\,\ref{Fig7}(e,f) marking $\Delta_{2}=0$; see also Fig.\,\ref{Fig1}(d).  
 
Below such magenta curve, where no JD appears in the trivial phase, a sizable diode's efficiency is confirmed revealing the period doubling effect due to MBSs. Above such magenta curve in the trivial phase where $\Delta_2 < 0$, however, a finite JD effect appears, which can be understood as a nontrivial effect of $\theta$ on the transparency dependence of the critical currents discussed before; here,  $\Delta_{2}$ vanishes in the trivial phase, a bunch of ABSs around zero energy affect the critical currents that then promote a nonzero JD efficiency. Thus, in the tunnelling regime and in the trivial phase, although the efficiency is small, it is preferred to have a negative $\Delta_{2}$ to get the diode effect. Notably, above the magenta curve, the outer gap is negative  $\Delta_{2}<0$ indicating that the lowest energy band has crossed the Fermi level, but still a bunch of zero energy states are present.
Nevertheless, below $\Delta_{2}=0$ no JD effect appears in the trivial phase. As a result, in the tunneling regime,    one can eliminate the effect of ABSs  and  JDs can  operate only with MBSs in the topological phase.

\section{Conclusions}
\label{sec:conclusion}
We   investigated  the emergence of the Josephson diode effect in Josephson junctions that can be realized using  superconductor-semiconductor hybrids under an homogeneous magnetic field. We have shown that the component of the Zeeman field parallel to the spin-orbit axis originates a phase-dependent energy spectrum that is asymmetric as a function of  the phase difference in the trivial and topological phases of transparent Josephson junctions with Andreev and Majorana bound states, respectively. We have then demonstrated that this asymmetry is the key ingredient for realizing current-phase curves with unequal positive and negative critical currents, which, signal non-reciprocal   transport that characterizes the Josephson diode effect.  We have further showed that the Josephson diodes can be controlled by the Zeeman field magnitude, where the diode's efficiency traces the gap closing and reopening as well as the oscillations of Majorana bound states in the topological phase. As a result, while the Josephson diode effect occurs in the presence of Andreev or Majorana bound states, it develops an intriguing response only in the topological phase due to the intrinsic spatial Majorana nonlocality that even enhances the Josephson diode's efficiency. Furthermore, we have discovered that, reducing the normal transmission also reduces the diode's efficiency in the trivial phase and  a Josephson diode effect only appears   in the topological phase with Majorana bound states. We have also verified that our findings remain robust at finite but low temperatures. Our work can therefore be useful for understanding the emergence of Josephson diodes under the presence of Andreev and Majorana bound states in semiconductor-superconductor hybrids as well as for identifying the topological phase.

 \begin{acknowledgments}
We thank L. Arrachea,   J.-F. Liu,  and W. Xu  for valuable discussions.  S. M. and J. C. acknowledge  financial support from the G\"{o}ran Gustafsson Foundation (Grant No. 2216), the Swedish Research Council  (Vetenskapsr\aa det Grant No.~2021-04121) and the Carl Trygger’s Foundation (Grant No. 22: 2093).  P.-H. Fu is supported by the National Natural Science Foundation of China (Grant No. 12174077).  The computations were enabled by resources provided by the National Academic Infrastructure for Supercomputing in Sweden (NAISS), partially funded by the Swedish Research Council through Grant Agreement No. 2022-06725. 
 \end{acknowledgments}

\appendix

\section{Josephson junctions with longer N regions and possible multichannel effects: spectrum and diode effect}\label{app:JD_with_long_N}
\begin{figure}[!t]
	\centering
	\includegraphics[width=\linewidth]{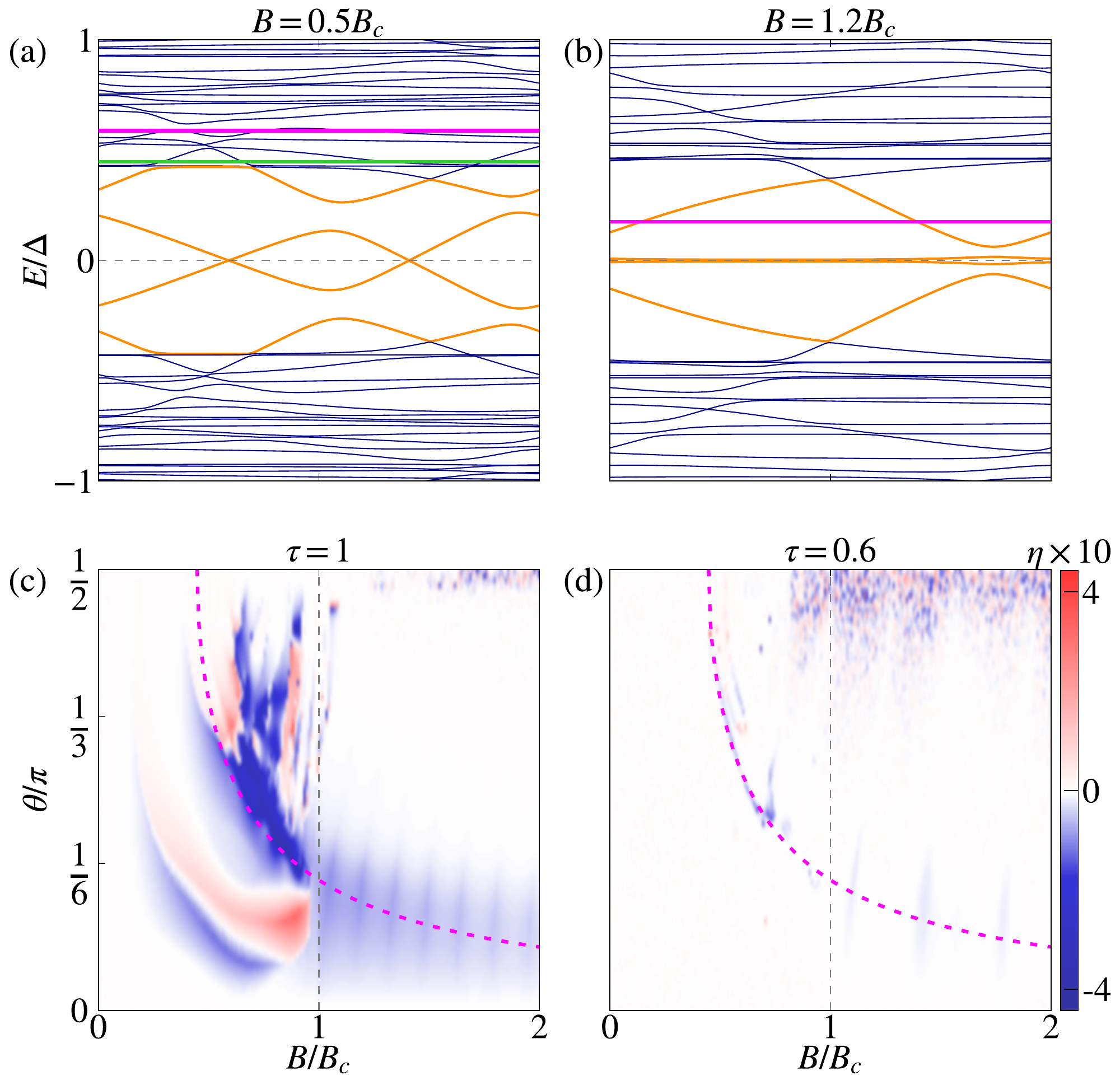}
	\caption{(a,b) Low-energy spectra for $B=0.5B_c$ and $B=1.2B_c$ with a long $\mathrm{N}$ of  $L_{\rm N} = 500\,\text{nm}$ .The superconducting regions are kept fixed at $L_\mathrm{S} = 2,{\rm \mu m}$. The green and magenta line represent the vanishing of $\Delta_{1}$ and $\Delta_{2}$ respectively. (c,d) Efficiency as a function of $B$ and $\theta$ in transparent and tunneling regime. Parameters: $L_\mathrm{S} = 2 {\rm \mu m}$, $L_{\rm N} = 500\,\text{nm}$, $\kappa_{\rm B}T = 0$, $\theta = \pi/12$ in (a,b), $\tau = 1$ in (a-c), $\tau = 0.6$ in (d), while the remaining parameters are the same as in Fig.~\ref{Fig1}.
	}
	\label{FigApp2}
\end{figure}

\begin{figure}[!t]
	\centering
	\includegraphics[width=\linewidth]{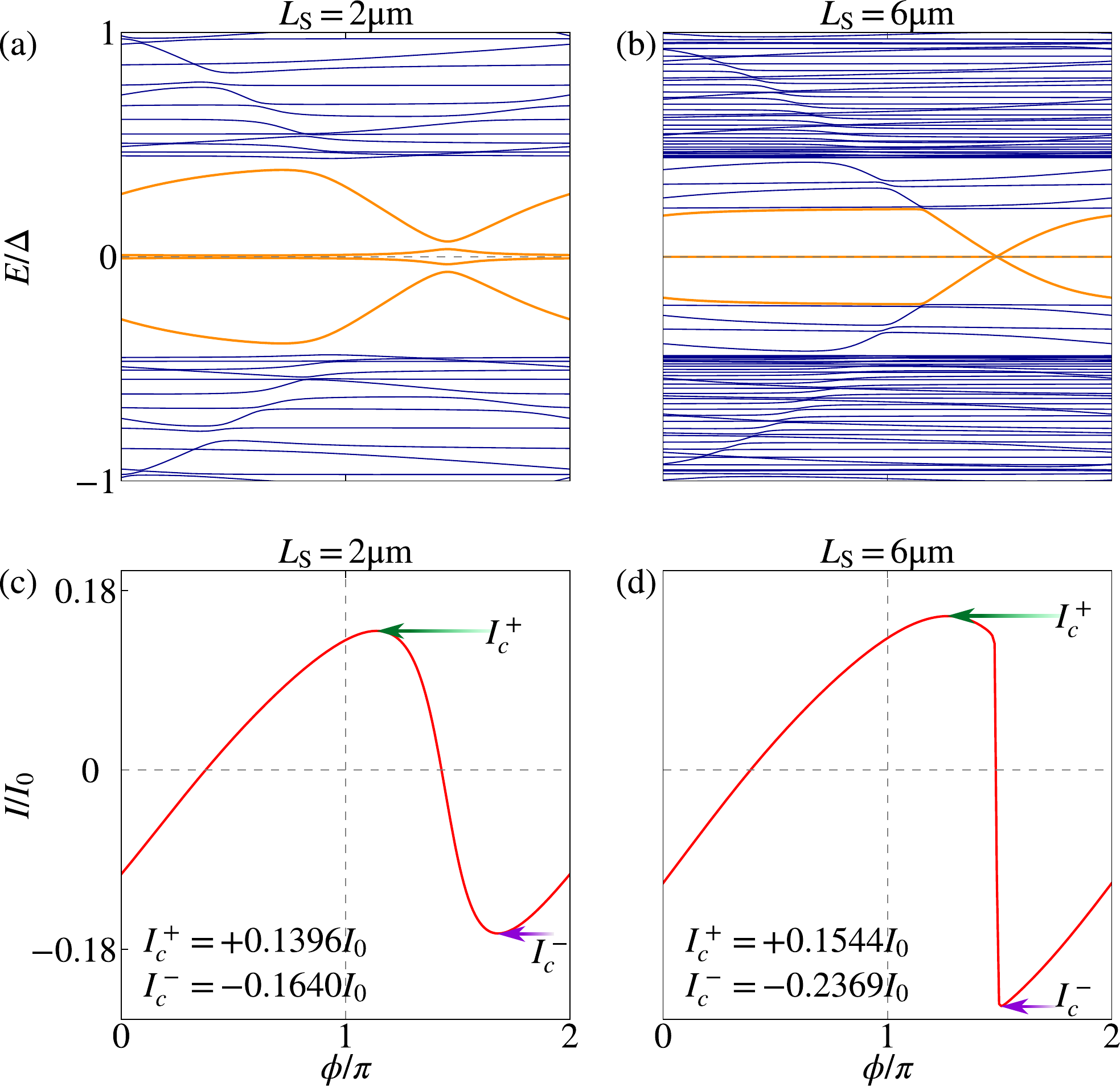}
	\caption{(a,b) Low-energy spectra for $L_\mathrm{S} = 2\mu\mathrm{m}$ and $L_\mathrm{S} = 6\mu\mathrm{m}$ with a fixed $L_\mathrm{N} = 20\mathrm{nm}$. The four energy states close to zero are shown in orange. (c,d) The supercurrent profile as a function of $\phi$ corresponding to $L_\mathrm{S} = 2\mu\mathrm{m}$ and $L_\mathrm{S} = 6\mu\mathrm{m}$. Parameters: $B = 1.2B_c$, $\theta = \pi/12$, $\kappa_{\rm B}T = 0$, $\tau = 1$, while the rest of the parameters are same as in Fig.~\ref{Fig2}.
	}
	\label{FigApp22}
\end{figure}

In the main text, the JD is realized using a short junction with a normal region $\mathrm{N}$ of length $L_\mathrm{N} = 20\mathrm{nm}$. Here we investigate a diode effect with a longer $\mathrm{N}$, specifically $L_\mathrm{N} = 500 \mathrm{nm}$. Increasing $L_\mathrm{N}$ has an immediate impact: more subgap states appear in both the trivial and topological phases \cite{cayao2018}; see Figs.\,\ref{FigApp2}(a,b). Similarly, extending the length of $\mathrm{N}$ pushes additional states above the superconducting gap. Due to the spin structure of the topological phase, however, the number of states is reduced compared to the trivial phase \cite{cayao2018}. These new states can acquire a strong dependence on the superconducting phase difference. For example, the four MBSs that occur at $\phi = 1.45\pi$ [see Fig.\,\ref{Fig2}(e) and Fig.\,\ref{FigApp22}(a)] in the short junction now shift to higher phase values; see Fig.\,\ref{FigApp2}(a). Such spectral changes affect the supercurrent \cite{cayao2018} and, consequently, the diode performance, with distinct signatures in the trivial and topological phases.

The diode efficiency $\eta$ as a function of the magnetic field $B$ and phase difference $\theta$ is presented in Figs.\,\ref{FigApp2}(c,d). 
In the trivial regime, $\eta$ is strongly modified due to contributions from ABSs and the quasicontinuum [Fig.\,\ref{FigApp2}(a)], whereas in the topological regime the efficiency is largely preserved because of the robustness of the spectrum [Fig.\,\ref{FigApp2}(b)]. Notably, in the trivial phase ($B < B_c$), the polarity reversal of $\eta$ no longer coincides with the $\Delta_{2} = 0$ curve observed in the short junction case. By contrast, in the topological regime, $\eta$ maintains its characteristic oscillatory profile, reflecting the presence of MBSs.

Interestingly, in the tunneling regime ($\tau = 0.6$), the diode efficiency $\eta$ is almost completely suppressed in the trivial phase, yet it preserves its oscillatory character in the topological phase with doubled periodicity, independent of $\theta$; see Fig.~\ref{FigApp2}(d). These findings highlight that while the trivial regime is highly sensitive to junction's transparency and normal region length, the MBS-induced diode effect in the topological phase remains a robust and distinctive signature.

Before ending this part, we would like to highlight that, similar to the effect of longer N regions, realistic systems might also have a finite thickness and hence multiple channels. Thus, multichannel JJs  might be hard to avoid under realistic circumstances and next we briefly point out some potential consequences in relation to the diode effect.  For a multichannel JJ, the behavior depends critically on the inter-subband coupling. In the simplest case where the subbands are not coupled, the system can be considered as $C_N$ independent single-channel nanowires. In this case, each subband can host MBSs in the topological phase, resulting in $C_N$ Majorana pairs. The overall diode efficiency is then expected to increase with the number of channels, since each channel contributes independently to the supercurrent. The critical field for the topological phase transition remains $B_c = \sqrt{\Delta^2 + \mu^2}$, and the dependence of the diode efficiency on $B$ and $\theta$ follows the same principles discussed earlier. However, if the subbands are coupled via a transverse Rashba spin-orbit coupling, e.g., $\alpha_{\rm R} \sigma_x \tau_0$, the situation changes qualitatively. In this case, pairs of MBSs from different channels can hybridize into finite-energy fermionic states. Specifically, for $C_N$ channels, $\left\lfloor C_N/2 \right\rfloor$ fermionic states emerge at finite energy, while the remaining $C_N \bmod 2$ MBSs survive as zero-energy states at each edge. The inter-subband coupling also changes the topological classification, leading to multiple topological phase transitions as discussed in Refs.~\cite{PhysRevB.84.144522, PhysRevX.7.021032, PhysRevLett.118.107701, jose2014_PRL}. This suggests that the diode efficiency spectrum in the multi-channel case with inter-subband coupling can be dramatically different and it deserves a proper and careful study.

\section{Low-energy spectrum and Josephson current with long superconductors}
\label{app:JC_with_long_S}
We previously discussed how increasing the superconductor's length enhances $\eta$. In this appendix, we present the corresponding low-energy spectra and Josephson current, which further demonstrate the improvement in $\eta$.

Figures \ref{FigApp22}(a,b) show the low-energy spectra for $L_\mathrm{S} = 2\mu\mathrm{m}$ and $L_\mathrm{S} = 6\mu\mathrm{m}$, respectively. Although Fig.\,\ref{FigApp22}(a) is identical to Fig.\,\ref{Fig2}(e),it is included here to allow direct comparison with the longer-superconductor case. The most immediate observation is that a longer $L_{\rm S}$ increases the number of discrete energy levels; see Fig.\,\ref{FigApp22}(b).
Moreover, when the superconducting regions are short ($L_{\rm S} < \xi_{\rm M}$, where $\xi_{\rm M}$ is the Majorana localization length), the overlap between Majoranas at opposite ends leads to a finite energy splitting near $\phi = 1.45\pi$, as seen in Fig.\,\ref{FigApp22}(a). However, when the superconducting regions are sufficiently long ($L_{\rm S} > \xi_{\rm M}$), the Majoranas become well-localized at the junction interfaces, resulting in exact zero-energy modes, as shown in Fig.\,\ref{FigApp22}(b).

This improved localization directly influences the supercurrent: both critical currents $I_c^+$ and $I_c^-$ increase with longer superconducting regions, as seen in Figs.\,\ref{FigApp22}(c,d). The enhancement occurs because better-localized Majoranas contribute more effectively to the Josephson current. As the Majoranas approach ideal zero modes, the diode efficiency rises [see the red curve in Fig.\,\ref{Fig6}(c)], owing to the more pronounced supercurrent asymmetry. Consequently, increasing $L_{\rm S}$ strengthens Majorana nonlocality and further enhances the Josephson diode effect, as discussed in Sec.\,\ref{sec:efficiency}.

\begin{figure}[!t]
	\centering
	\includegraphics[width=\linewidth]{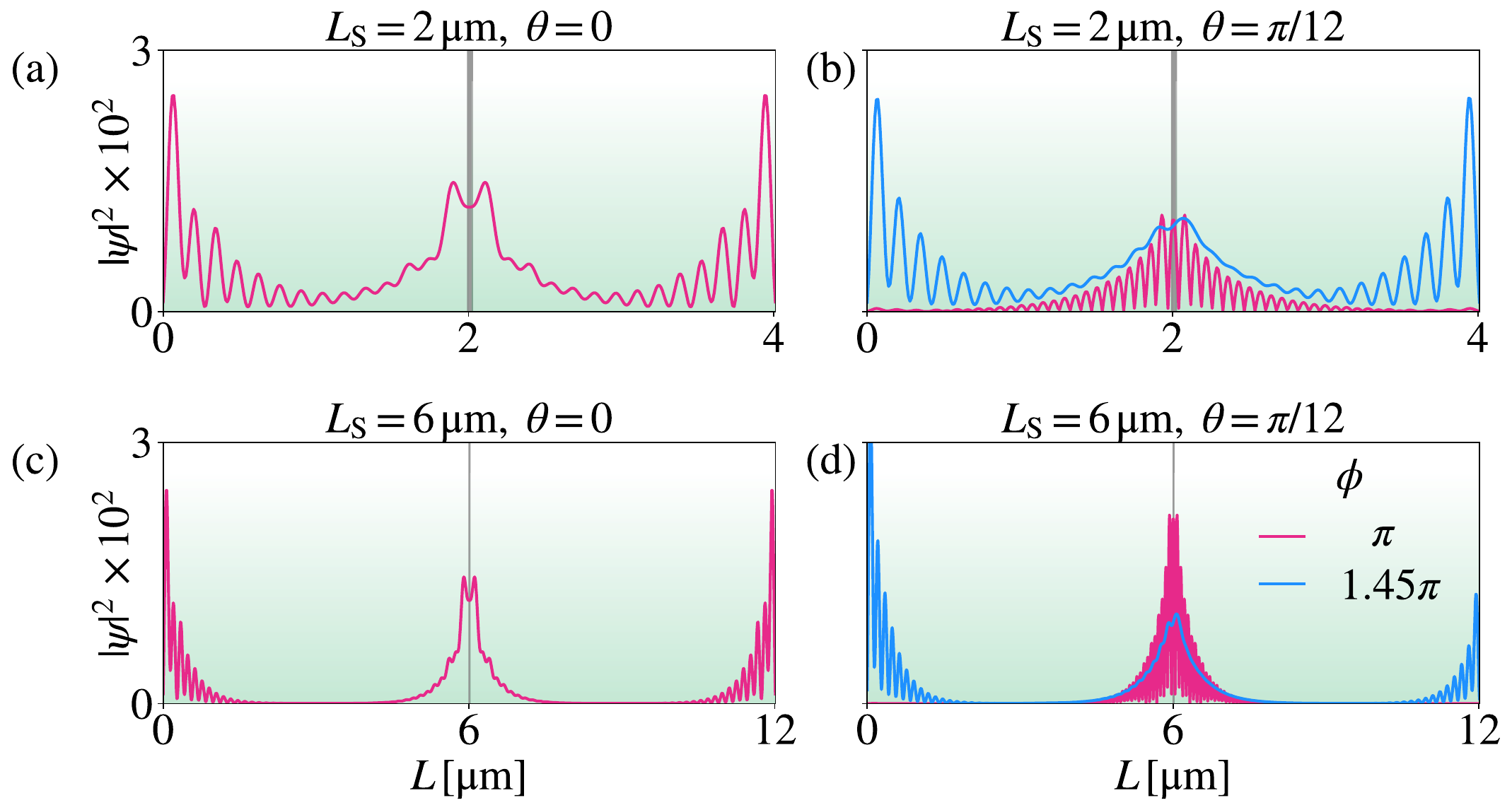}
	\caption{The distribution of the modulus squared of the wave function, $|\psi|^2$, is shown for different values of $L_\mathrm{S}$ and $\theta$. The narrow gray area represents the normal region, which is sandwiched between two superconductors depicted in green. The curves in the blue and red correspond to $\phi = \pi$ and $\phi = 1.45 \pi$ respectively. Parameters: $L_{\rm N} = 20\,\text{nm}$, $\tau = 1$, $B = 1.2 B_c$, while the remaining parameters are the same as in Fig.~\ref{Fig1}.
	}
	\label{FigApp3}
\end{figure}

\section{Majorana wavefunctions for distinct $\theta$ in short Josephson junctions}\label{app:Localization_MBS_in_JD}
In Sec.\,\ref{sec:abs}, we described the splitting of MBSs occurring either at $\phi = \pi$ or at $\phi = 1.45\pi$ depending on the value of $\theta$. Here, we clarify the origin of this splitting by examining the spatial localization of the MBSs.

The distribution of the modulus squared of wave function, $|\psi|^2$ of the MBSs is shown in Fig.\ref{FigApp3}. In each panel, the narrow gray region represents $\mathrm{N}$ separating the $\mathrm{S_L}$ and $\mathrm{S_R}$, which are indicated in light green. As discussed in the main text, at $\phi = \pi$ and $\theta = 0$, four Majorana modes appear [see Fig.\,\ref{Fig2}(d)], localized at the ends of the superconductors as illustrated in Fig.\ref{FigApp3}(a).
When $\theta = \pi/12$ and $\phi = \pi$, the Majoranas are localized at the inner ends of the superconductors, as shown in Fig.\ref{FigApp3}(b), where two Majoranas are present. In addition, two more Majorana modes appear at the outer ends of the superconductors at $\phi = 1.45\pi$ [see Fig.\,\ref{Fig2}(e)], as indicated by the blue curve in Fig.\ref{FigApp3}.

Figures\,\ref{FigApp3}(a,b) correspond to $L_\mathrm{S} = 2\mu \mathrm{m}$, which is smaller than twice the Majorana localization length, that is, $L_\mathrm{S} < 2\xi_\mathrm{M}$. This results in a finite overlap between the wave functions of the left and right Majoranas within each superconducting segment, as evident from Figs.\,\ref{FigApp3}(a,b). In contrast, when the superconductors are longer such that $L_\mathrm{S} > 2\xi_\mathrm{M}$, the wave functions of the Majoranas do not overlap, as shown in Figs.\,\ref{FigApp3}(c,d). In Fig.\,\ref{FigApp3}(d), the amplitudes of the wave functions at the left and right ends of both superconductors differ slightly.

These observations confirm that a finite $\theta$ does not destroy the MBSs in the topological regime. Instead, the modes remain well-localized at the superconductor boundaries, consistent with the description in Sec.\,\ref{sec:abs}.

\section{Effect of temperature on the  efficiency of the Josephson diodes}
\label{app:Efficiency_temperature}
In the main text, we have discussed efficiency $\eta$ both for a transparent JJ and a JJ in the tunnelling regime. In the first case, we have observed the finite $\eta$ when the system hosts ABSs and MBSs. In contrast, in the tunneling regime, $\eta$ exhibits a strong dependence on MBSs alone. Moreover, in this regime, $\eta$ shows a re-entrant behavior at $B = B_c$ and an oscillatory pattern with double periodicity in the nontrivial phase. 
However, the effect of a small but finite temperature differs from that in the tunneling regime. Although a slight increase in temperature suppresses the MBS-driven oscillations of efficiency in the topological phase, but the efficiency remains robust. We have verified that current phase curve follows a nonreciprocal behavior in presence of such small temperature. We only show the critical currents and the efficiency below.

\begin{figure}[!t]
	\centering
	\includegraphics[width=\linewidth]{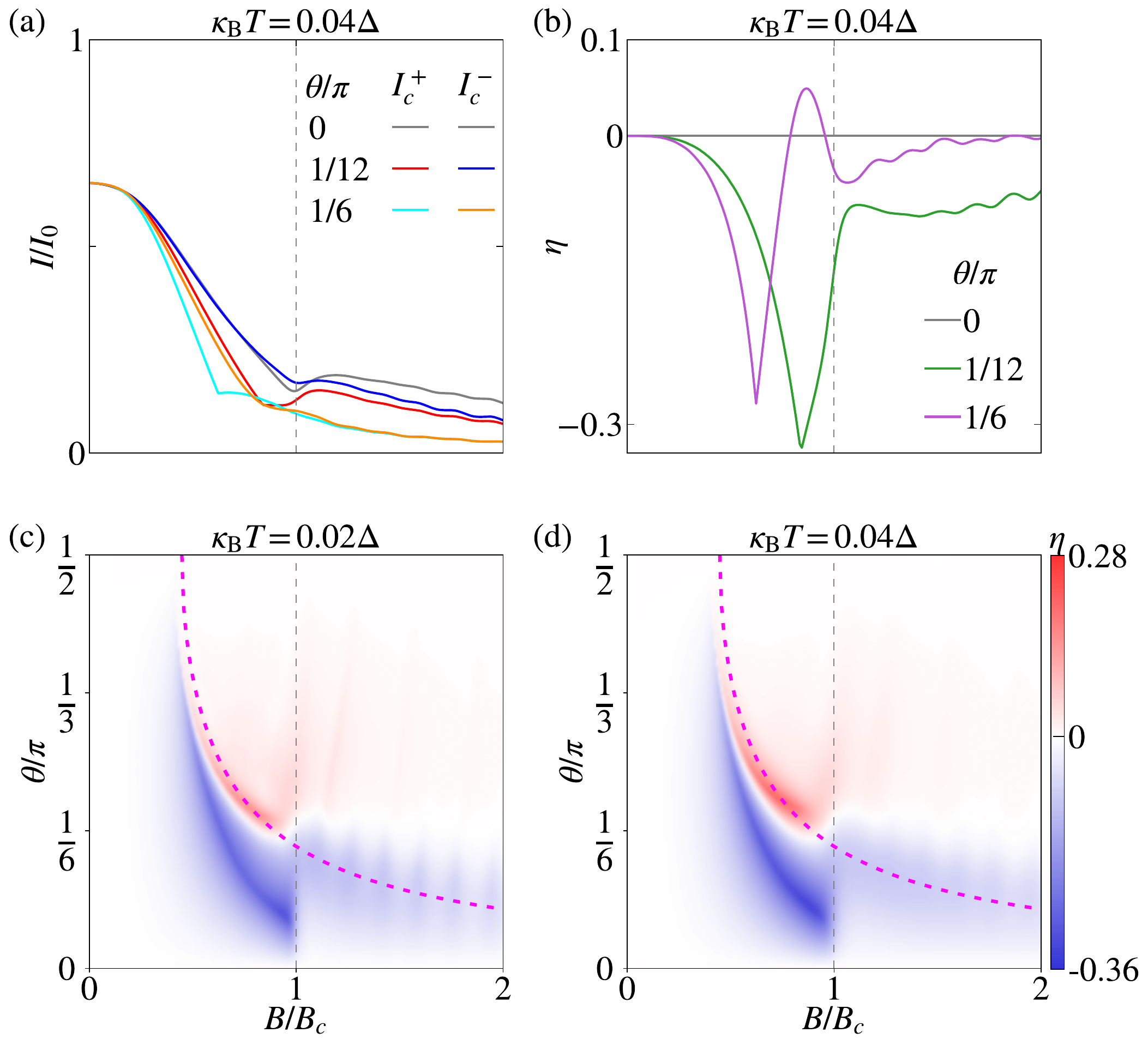}
	\caption{(a) The critical currents and (b) the efficiencies are shown for $k_BT = 0.04\Delta$ as a function $B/B_c$. Panels [(c) and (d)] displays the efficiency as a function of $B$ and $\theta$, where the magenta dotted line represents the $\Delta_2 = 0$ curve. The vertical dashed line in each panel represents the vanishing of bulk gap, $\Delta_{1} =0$.}
	\label{FigApp}
\end{figure}	

To provide a complete picture of the effect of a small temperature, we present the variation of both the critical currents $I_c^\pm$  and the efficiency $\eta$ with the Zeeman field $B$ in Fig.~\ref{FigApp}(a,b) for three representative values of $\theta$. For the calculation, we set a small temperature $\kappa_{\rm B}T = 0.04\Delta$, 
ensuring it remains below the superconducting order parameter.
Furthermore, we present $\eta$ as a function of both $B$ and $\theta$ in Fig.\,\ref{FigApp}(c,d) for two different temperatures $\kappa_{\rm B}T = 0.02\Delta$ and $\kappa_{\rm B}T = 0.04\Delta$.
For $\theta = 0$, the identical critical currents, $I_c^\pm$, exhibit features [see Fig.~\ref{FigApp}(a)] similar to those observed at zero temperature in the $B < B_c$ region. 
At $B = B_c$, a kink-like feature appears. In the $B > B_c$ region, an oscillatory pattern emerges but with a suppressed amplitude. For $\theta \neq 0$, the distinct critical currents, $I_c^+$ and $I_c^-$, display a larger difference in the trivial regime compared to their zero-temperature counterparts, see the red (cyan) and blue (yellow) curves for $\theta =\pi/12$ ($\pi/6$) in Fig.\,\ref{FigApp}(a).  However, in the topological regime, the Majorana-driven oscillatory pattern is now suppressed, similar to the case of $\theta = 0$. Thus, the small temperature primarily smooths out oscillations and suppresses features related to MBSs, while enhancing the difference between critical currents in the trivial regime.

When it comes to the efficiency at the non-zero temperature, $\eta$ exhibits enhanced values in the trivial regime as compared to that for $\kappa_{\rm B}T = 0$, see the purple and green curves in Fig.\,\ref{FigApp}(b). This enhancement arises from the increased difference between $I_c^+$ and $I_c^-$ in the trivial regime. In the topological phase, $\eta$ retains its oscillatory pattern but with a reduced amplitude. Moreover, the presence of a nonzero temperature smooths the curve, making the peaks and dips less pronounced.
Further insights on $\eta$ is obtained from Fig.\,\ref{FigApp}(c,d), where we illustrate its dependence on both $B$ and $\theta$ for different values of $\kappa_{\rm B}T$. In the trivial regime, both positive and negative $\eta$ exhibit increased magnitudes.
Additionally, as $\theta$ increases, the maximum $\eta$ shifts away from $B = B_c$. Notably, the locus of $\eta = 0$ (represented by the thin white region sandwiched between the red and blue regions in the trivial regime) moves away from the $\Delta_2 = 0$ curve. This contrasts with the $\kappa_{\rm B}T = 0$ case, where both coincide, as shown in Figs.\,\ref{Fig6}(a,b). Importantly, in the topological phase, the amplitude of the MBS-driven oscillatory pattern diminishes as the temperature increases. 
Therefore, the non-vanishing nature of the efficiency suggests that it remains robust against a small but finite temperature.


\bibliography{biblio}

\end{document}